\documentclass[preprint2,numberedappendix]{emulateapj}

\newcommand{\Halpha}{\ensuremath{{\rm H \alpha} }}
\newcommand{\Hbeta}{\ensuremath{{\rm H \beta} }}

\newcommand{\OIIIa}{\ensuremath{{\rm [OIII] \lambda 4959} }}
\newcommand{\OIIIb}{\ensuremath{{\rm [OIII] \lambda 5007} }}
\newcommand{\OII}{\ensuremath{{\rm [OII] \lambda 3727,3729} }}
\newcommand{\NeIII}{\ensuremath{{\rm [NeIII] \lambda 3870} }}
\newcommand{\NII}{\ensuremath{{\rm [NII] \lambda 6584} }}

\newcommand{\p}{\ensuremath{\pm \;}}
\newcommand{\Msun}{\ensuremath{M_{\odot}}}
\newcommand{\M}{\ensuremath{M}}
\newcommand{\HST}{\emph{HST}}
\newcommand{\microm}{\ensuremath{{\rm \mu m } }}
\newcommand{\logOH}{\ensuremath{\log(\mathrm{O/H})}}

\newcommand{\Align}{\phantom{2}}

\newcommand{\alignnn}{\phantom{22.2 }}
\newcommand{\alignp}{\phantom{2. }}
\newcommand{\note}{\tablenotemark{$\ast$}}

\usepackage{graphicx}
\usepackage{amsmath}

\shorttitle{The Fundamental Metallicity Relation at High Redshift}
\shortauthors{Belli et al.}
\slugcomment{Accepted for publication in the Astrophysical Journal}

\begin{document}

\title{Testing the Universality of the Fundamental Metallicity Relation at High Redshift Using Low-Mass Gravitationally Lensed Galaxies}

\author{Sirio Belli\altaffilmark{1}, Tucker Jones\altaffilmark{2}, Richard S. Ellis\altaffilmark{1}, Johan Richard\altaffilmark{3}}
\altaffiltext{1}{Department of Astronomy, California Institute of Technology, MS 249-17, Pasadena, CA 91125, USA}
\altaffiltext{2}{Department of Physics, University of California, Santa Barbara, CA 93106, USA}
\altaffiltext{3}{Centre de Recherche Astrophysique de Lyon, Universit\'e Lyon 1, 9 Avenue Charles Andr\'e, 69561 Saint Genis Laval Cedex, France}

\begin{abstract}

We present rest-frame optical spectra for a sample of 9 low-mass star-forming galaxies in the redshift range $1.5 < z < 3$ which are gravitationally lensed by foreground clusters. We used Triplespec, an echelle spectrograph at the Palomar 200-inch telescope that is very effective for this purpose as it samples the entire near-infrared spectrum simultaneously. By measuring the flux of nebular emission lines we derive gas phase metallicities and star formation rates, and by fitting the optical to infrared spectral energy distributions we obtain stellar masses. Taking advantage of the high magnification due to strong lensing we are able to probe the physical properties of galaxies with stellar masses in the range $ 7.8 < \log \M/\Msun < 9.4 $ whose star formation rates are similar to those of typical star-forming galaxies in the local universe. We compare our results with the locally determined relation between stellar mass, gas metallicity and star formation rate. Our data are in excellent agreement with this relation, with an average offset $ \langle \Delta \logOH \rangle = 0.01 \p 0.08 $, suggesting a universal relationship. Remarkably, the scatter around the fundamental metallicity relation is only 0.24 dex, smaller than that observed locally at the same stellar masses, which may provide an important additional constraint for galaxy evolution models.

\end{abstract}



\section{Introduction}

The gas-phase metallicity represents a fundamental property of galaxies and can be used to investigate the complex physical processes that govern galaxy evolution. It mainly traces the star formation history, as metals produced in stars are ejected into the interstellar medium (ISM), but the exchange of material between the galaxy and the intergalactic medium (IGM) also plays an important role. The accretion of metal-poor gas from the IGM can dilute the metal content of the gas in a galaxy. Also, stellar winds can substantially lower the metallicity by ejecting metals.

Despite the complexity of these processes, a clear relation between galaxy luminosity and metallicity has been known since the work of \citet{lequeux79}. Recently, thanks to the vast amount of spectroscopic and photometric data available from the Sloan Digital Sky Survey (SDSS), it has become clear that the physical parameter that correlates most strongly with metallicity is the galaxy stellar mass \citep{tremonti04}. This mass-metallicity relation, in which galaxies of higher masses contain larger metallicities, is remarkably tight over 3 orders of magnitude in stellar mass, with a dispersion of only 0.10 dex in metallicity.

A natural explanation for the observed mass-metallicity relation is the outflow of metal-enriched gas driven by star formation. Because of the lower gravitational potential, low-mass galaxies lose a higher fraction of their gas, with a consequent decrease in metallicity \citep{larson74, garnett02, tremonti04}. An alternative possibility is that lower mass galaxies are less metal-rich because their star formation history has been developed more gradually \citep{ellison08}, in agreement with the now-familiar effect of \emph{downsizing} \citep{cowie96}.

Different models of galaxy formation and evolution are able to match the mass-metallicity relation in the local universe, but have dissimilar predictions for high redshift galaxies \citep[e.g.][]{delucia04, dave07, tassis08, dave11II, yates12}. Observing the redshift evolution of the mass-metallicity relation can therefore differentiate these models. Observations at different redshifts have shown a clear evolution with cosmic time, with lower metallicity at higher redshift, for a fixed mass \citep{savaglio05, erb06a, maiolino08, mannucci09, zahid11, yuan13}.

However, it is important to recognize that high-redshift studies target galaxy populations that are different from those found typically in the local Universe. The evolution of the mass-metallicity relation could then be the result of a selection effect rather than a change in the physical properties of the galaxies with cosmic time. 

Among the differences between local and high-redshift galaxies, star formation activity is one of the most important. At earlier cosmic times, the star formation rate (SFR) was on average much higher than today, because galaxies contained a larger amount of cold gas. Also, most of the high-redshift surveys are magnitude-limited in the rest-frame UV, and therefore tend to select galaxies with high SFR. The combination of these two effects makes it very difficult to compare the metallicity of galaxies at different redshifts with the same stellar mass and star formation rate. It is then essential to study the relation between SFR and metallicity, since this could have important consequences on the interpretation of the observed evolution of the mass-metallicity relation.
 
In fact, \citet{mannucci10} found that the local mass-metallicity relation is different for samples of galaxies with different star formation rates. Furthermore, they showed that the SDSS galaxies lie on a tight 3D surface in the mass-metallicity-SFR space, with a dispersion of only 0.053 dex in metallicity \citep[see also][]{lara-lopez10}. According to this \emph{fundamental metallicity relation} (FMR), at fixed stellar mass SFR and metallicity are anti-correlated. If this relation holds independently of cosmic time, a galaxy population at high redshift will tend to have a low average metallicity because of its high SFR. In this scenario the evolution of the mass-metallicity relation is driven by the shifting of galaxy populations on the SFR-mass plane, rather than being directly caused by the evolution of some physical process. Clearly, testing the fundamental metallicity relation at different redshifts is of primary importance.

The redshift evolution of the FMR was first explored by \citet{mannucci10} using samples from the literature, and they concluded that the local relation is a good fit for any star-forming galaxy up to $z \sim 2.2$. But high-redshift observations are biased towards high-SFR galaxies, and a direct test using galaxies with the same range of star formation rates that is seen in the SDSS sample (SFR $<$ 10 \Msun/yr) is still lacking. Additional difficulties come from the fact that to measure the metallicity one needs the rest-frame optical emission lines, which at $z \sim 2$ are redshifted into the near-infrared, a spectral region where sky emission is strong. 

One way to probe lower star formation rates with the current technology is to take advantage of strong gravitational lensing. The magnification induced by foreground galaxy clusters allows one to reach faint objects, corresponding to stellar masses and SFRs (on average) lower than the values achievable without lensing. Recent studies of the FMR for high-redshift lensed galaxies found a general agreement with the local relation, although with a very large scatter \citep{richard11, wuyts12, christensen12}. 

Testing the universality of the fundamental metallicity relation is important not only for understanding the evolution of the mass-metallicity relation, but also to constrain models of galaxy evolution. For example, \citet{dave11II} consider a simple model in which inflow, outflow and star formation are in equilibrium and determine the gas metallicity. This scenario can qualitatively explain the dependence of the mass-metallicity relation on the star formation rate: at a fixed stellar mass a high SFR is caused by a large inflow, that in turn implies a low metallicity. If a galaxy is perturbed, e.g.\ by a merger, it will move away from the FMR and, after some time, will return to the equilibrium configuration. This equilibrium timescale determines the scatter in the mass-metallicity-SFR relation. So long as the yield of metals per unit star formation and the mass loading factor (i.e.\ the ratio between outflow and star formation) are constant, the equilibrium relation implies a universal FMR independent of redshift. Although this and other simple analytic models \citep{dayal12, dave12} succeed in explaining the local fundamental metallicity relation, we are still far from a detailed understanding of the relevant physical processes. High-redshift observations are essential for quantitative tests of hydrodynamical simulations of galaxy evolution.

In this work we study a sample of low-luminosity, $z \sim 2$ gravitational arcs with magnification factors of ${\sim} 10$--100. We used the Triplespec spectrograph on the Palomar 200-inch telescope that features a good sensitivity and covers the full near-infrared wavelength range. This characteristic makes it an ideal instrument for such a study, because it allows us to observe all diagnostic lines of interest simultaneously. In addition to the efficiency of observation, a particular benefit is the ability to measure emission line ratios, the diagnostics of gas-phase metallicities, in a single exposure, mitigating uncertainties that arise from variations in weather conditions. Also, we mainly rely on the emission lines \OIIIb, \OII, \Halpha\ and \Hbeta\ for measuring the metallicity, so that we are not limited by the requirement of detecting the faint \NII\ line. For these reasons we probe stellar masses and star formation rates that are on average lower than the ones of previously studied samples of lensed galaxies.

The sample of gravitational arcs, the spectroscopic observations, and data reduction are described in Section 2. In Section 3 we present the photometric measurements and the fitting of the spectral energy distribution, while in Section 4 we explore the galaxy physical properties using the measured line fluxes. We discuss the constraints of these measurements on the evolution of the fundamental metallicity relation in Section 5, and the implications for galaxy evolution models in Section 6. We assume a $\Lambda$ cold dark matter ($\Lambda$CDM) cosmology with $\Omega_\Lambda = 0.7$,  $\Omega_\mathrm{m} = 0.3$ and $H_0 = 70 \mathrm{ \ km\ s^{-1} Mpc^{-1} } $. Magnitudes are given in the AB system.

\begin{deluxetable*}{llllccc}
\tabletypesize{\footnotesize}
\tablewidth{0pc}
\tablecaption{Sample of Gravitational Arcs \label{tab:sample}}
\tablehead{
\colhead {Cluster} & \colhead{Arc}  & \colhead{$\mu$\tablenotemark{a}} & \colhead{Reference} & \colhead{Run} & \colhead{Exp. Time} & \colhead{Redshift\tablenotemark{b}} }
\startdata
  A611   	& 1.2, 1.3	& 19.6 \p 3.0			& \citet{newman09}\tablenotemark{c}	& B 	& 1h20min		& 1.4902 		\\
  RXJ2129	& 1.1, 1.2	& \phantom{2.}61 \p 17 		& \citet{richard10}			& A 	& 5h30min		& 1.5221 		\\
  A1413 	& 2.1, 2.2	& 23.9 \p 6.4			& \citet{richard10}			& B, D	& 3h00min		& 2.0376 		\\
  A1835 	& 7.1		& \phantom{2.}88 \p 30		& \citet{richard10}			& B 	& 1h45min		& 2.0733 		\\
  RXJ1720 	& 1.1, 1.2	& 22.5 \p 9.1			& \citet{richard10}			& A	& 4h00min		& 2.2200		\\
  A773		& 1.1, 1.2	& 27.9 \p 7.9			& \citet{richard10}			& C	& 3h15min		& 2.3032		\\
  MACS0717	& 13.1		& \phantom{2}7.2 \p 3.0		& \citet{limousin11}			& C, D 	& 5h30min		& 2.5515		\\
  A383		& 3C, 4C	& 23.9 \p 3.3			& \citet{newman11}			& C  	& 2h00min		& 2.5771		\\
  A1689		& 1.1, 1.2	& \phantom{2.}57 \p 23		& \citet{coe10}				& D  	& 6h30min		& 3.0421		\\
  A1703 	& 3.1, 3.2	& \phantom{2.}46 \p 20		& \citet{richard09}			& B  	& 3h30min		& 3.2847 		
\enddata
\tablenotetext{a}{Gravitational magnification. For multiply imaged sources, this is the sum of the magnifications.}
\tablenotetext{b}{Redshifts measured from \OIIIb, except for RXJ1720, for which \Halpha\ was used. The uncertainties are always less than 0.0003.}
\tablenotetext{c}{Magnification factor calculated after updating the lensing map with the new arc redshift, see Section \ref{sec:spectra}.}
\end{deluxetable*}

\begin{deluxetable}{clc}
\tabletypesize{\footnotesize}
\tablewidth{0pc}
\tablecaption{Observing Runs \label{tab:runs}}
\tablehead{
\colhead{Run} & \colhead{Date} & \colhead{Seeing (arcsec)} }
\startdata
 A   		& 2010 August 23, 25, 26	&	0.9 - 1.3	\\
 B   		& 2011 April 10, 11, 12, 13 	&	0.9 - 1.3	\\
 C   		& 2012 January 12, 13, 14  	&	1.1 - 2.0	\\
 D   		& 2012 April 29, 30, May 1  	&	0.9 - 1.5
\enddata
\tablecomments{The seeing was measured in the $K_S$ band.}
\end{deluxetable}


\section{Data}

\subsection{Sample Selection}
\label{sec:data1}

We selected our sample of gravitational arcs from the literature according to the following three criteria.

First, we considered only galaxy clusters with a well-constrained lens model. This allowed us to select arcs of known magnification $\mu \gtrsim 10$.

Second, the observed (i.e.\ not corrected for lensing) arc magnitude must be $R \lesssim 23 $, so that observations with Triplespec at Palomar are feasible. This means that we can probe intrinsic magnitudes $R \gtrsim 25.5$, fainter than the limits of typical non-lensing surveys.

Third, the arc should have a known spectroscopic redshift such that the emission lines from [OII] to \Halpha\ fall in the wavelength range observable with Triplespec. The ideal range is $2 < z < 2.5$, but we can measure metallicities and star formation rates using lines available for galaxies from $z \sim 1.5$ to $z \sim 3$.

This selection provides us with 10 sources viewed through 10 distinct clusters (see Table~\ref{tab:sample}). For the arc in RXJ1720 only \Halpha\ is observable because the other diagnostic lines are unfortunately obscured by night sky emission. We do not attempt any analysis on this object, but we include it in Table~\ref{tab:sample} because ours is the first redshift measurement for this arc obtained from a rest-frame optical emission line. 

From now on we will refer to the gravitational arcs by the names of the corresponding galaxy clusters, e.g.\ A1835 for A1835 arc 7.1.

\subsection{Spectroscopy}
\label{sec:spectroscopy}

All spectroscopic data were taken with Triplespec on the 200-inch Hale Telescope at Palomar Observatory over the course of four observing runs (see Table~\ref{tab:runs}). Triplespec is a near-infrared cross-dispersed spectrograph that simultaneously covers the wavelength range 1--2.4 \microm\ with a resolution R $\sim 2700$ \citep{herter08}.

The $1 \times 30$ arcsec long slit was positioned on the targets as shown in Fig. \ref{fig:centerpiece} via a blind offset from a bright star. For each target we typically undertook many exposures of 300--450 seconds each, using a two-point dithering pattern. The position of the target along the slit and the dithering offset were carefully chosen for each arc, avoiding any overlap of the arc with foreground cluster galaxies, and leaving enough blank sky along the slit to reliably measure and subtract background emission. When it was possible to arrange a multiply imaged system in the slit, the spectra, if resolved, were reduced and extracted separately and then combined.

The spectroscopic data were reduced using a modified version of \emph{Spextool} \citep{cushing04, vacca04}. For each target we extracted the spectrum from each A$-$B pair and then combined the 1D spectra. The aperture for the boxcar extraction was defined using the [OIII] emission profile in the stack of many A$-$B pairs, since the line emission is rarely detected in single frames. 

Flux calibration and correction for telluric absorption were performed using \citet{elias82} A-type standard stars. Note that this procedure also corrects for the variation of effective seeing with wavelength. Although the absolute flux calibration can be very uncertain, it affects only the emission line fluxes but not their ratios, which are used in calculating gas-phase metallicities. This is one of the main advantages of using Triplespec, which allows one to observe the entire near-infrared spectrum at once. However, SFR measurements are affected by absolute flux calibration, that therefore needs to be carefully quantified.

\begin{figure*}[htbp]

\begin{minipage}{\textwidth}
   \centering
 \raisebox{-0.5\height}{\includegraphics[width=0.15\textwidth]{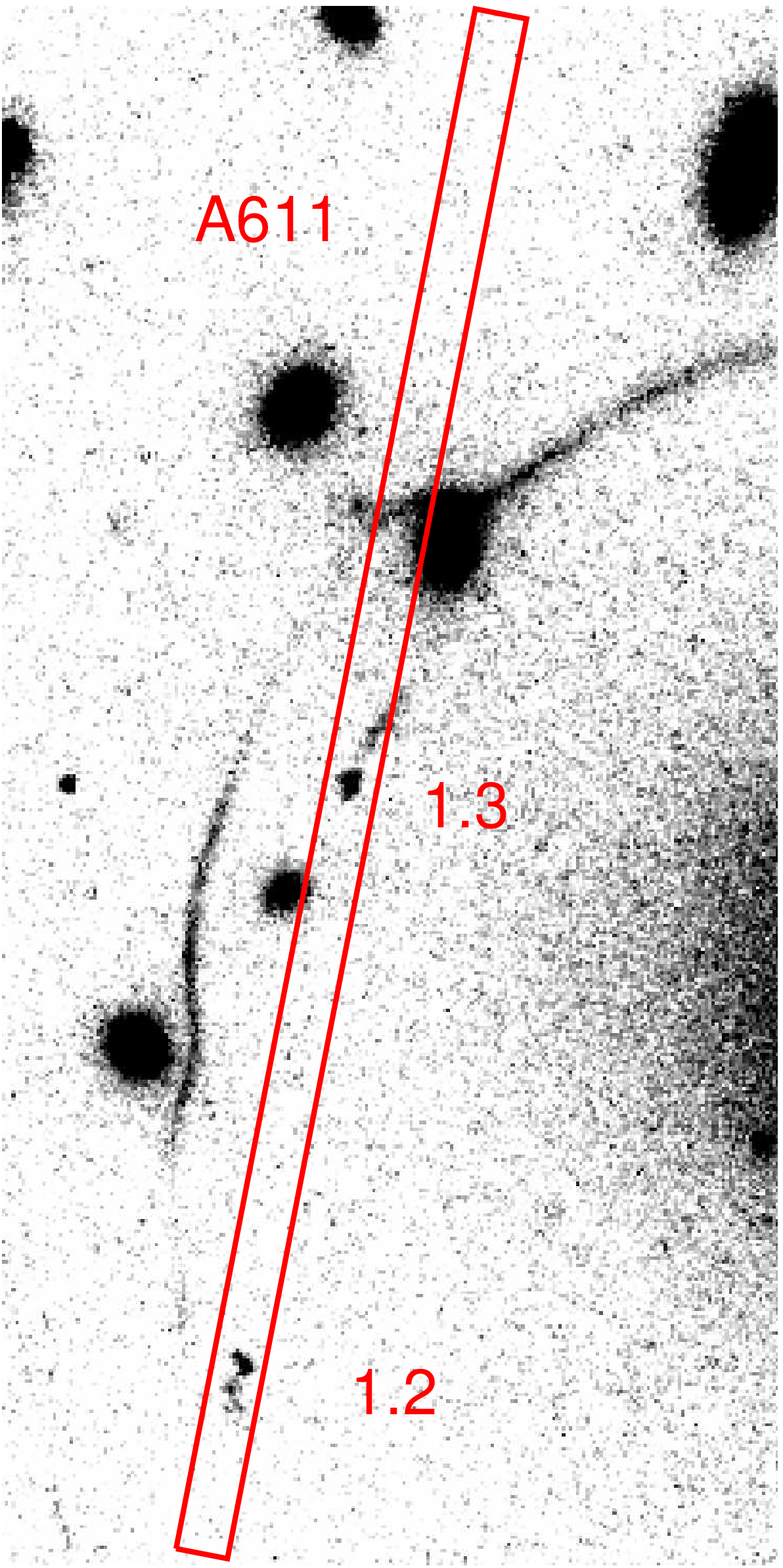}}
   \hspace*{0.02\textwidth}
 \raisebox{-0.5\height}{\includegraphics[width=0.32\textwidth]{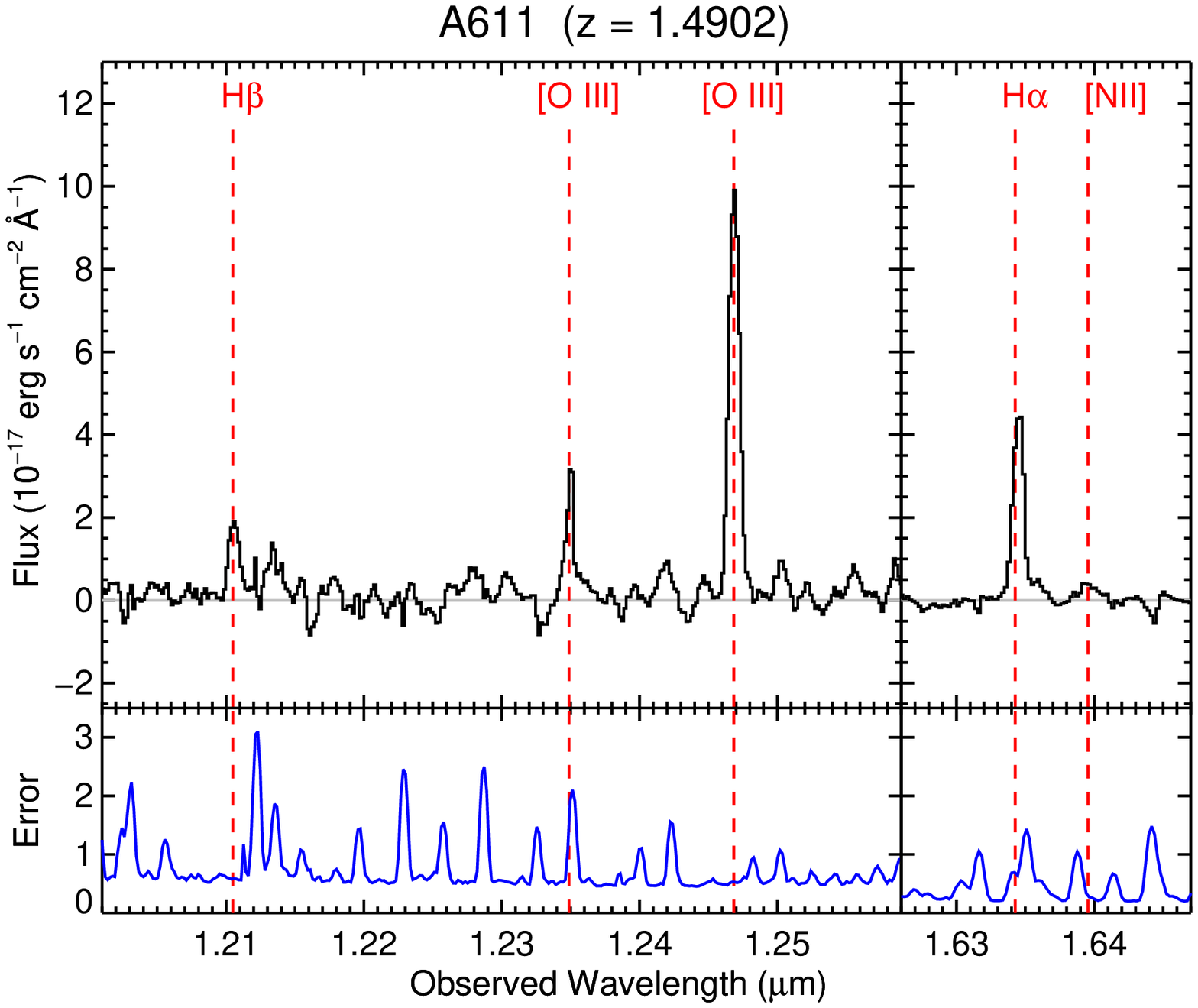}}
   \hspace*{0.02\textwidth}
 \raisebox{-0.5\height}{\includegraphics[width=0.32\textwidth]{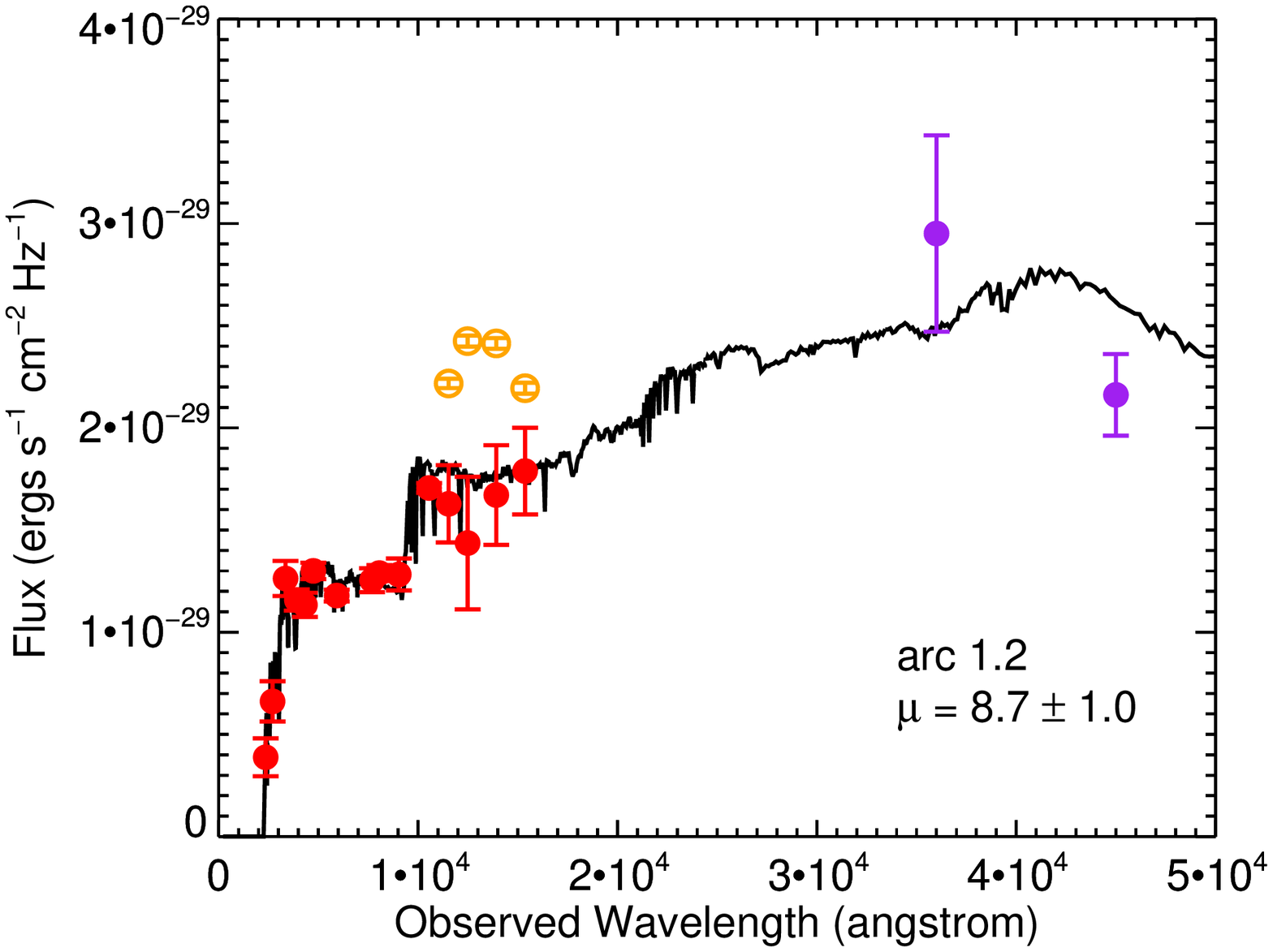}}
\end{minipage}
  
  \vspace*{0.02\textwidth}
  
\begin{minipage}{\textwidth}
   \centering
 \raisebox{-0.5\height}{\includegraphics[width=0.15\textwidth]{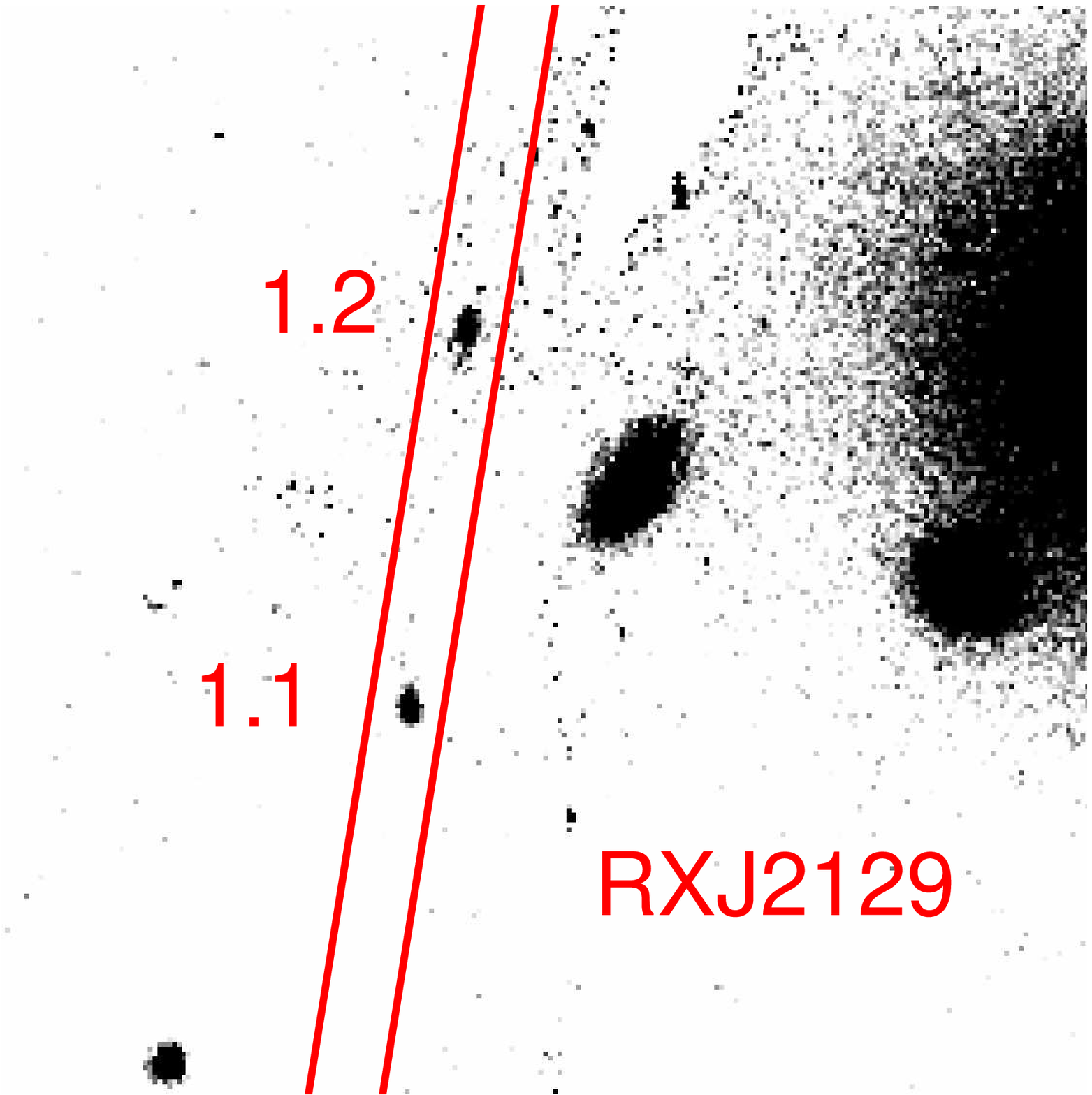}}
   \hspace*{0.02\textwidth}
 \raisebox{-0.5\height}{\includegraphics[width=0.32\textwidth]{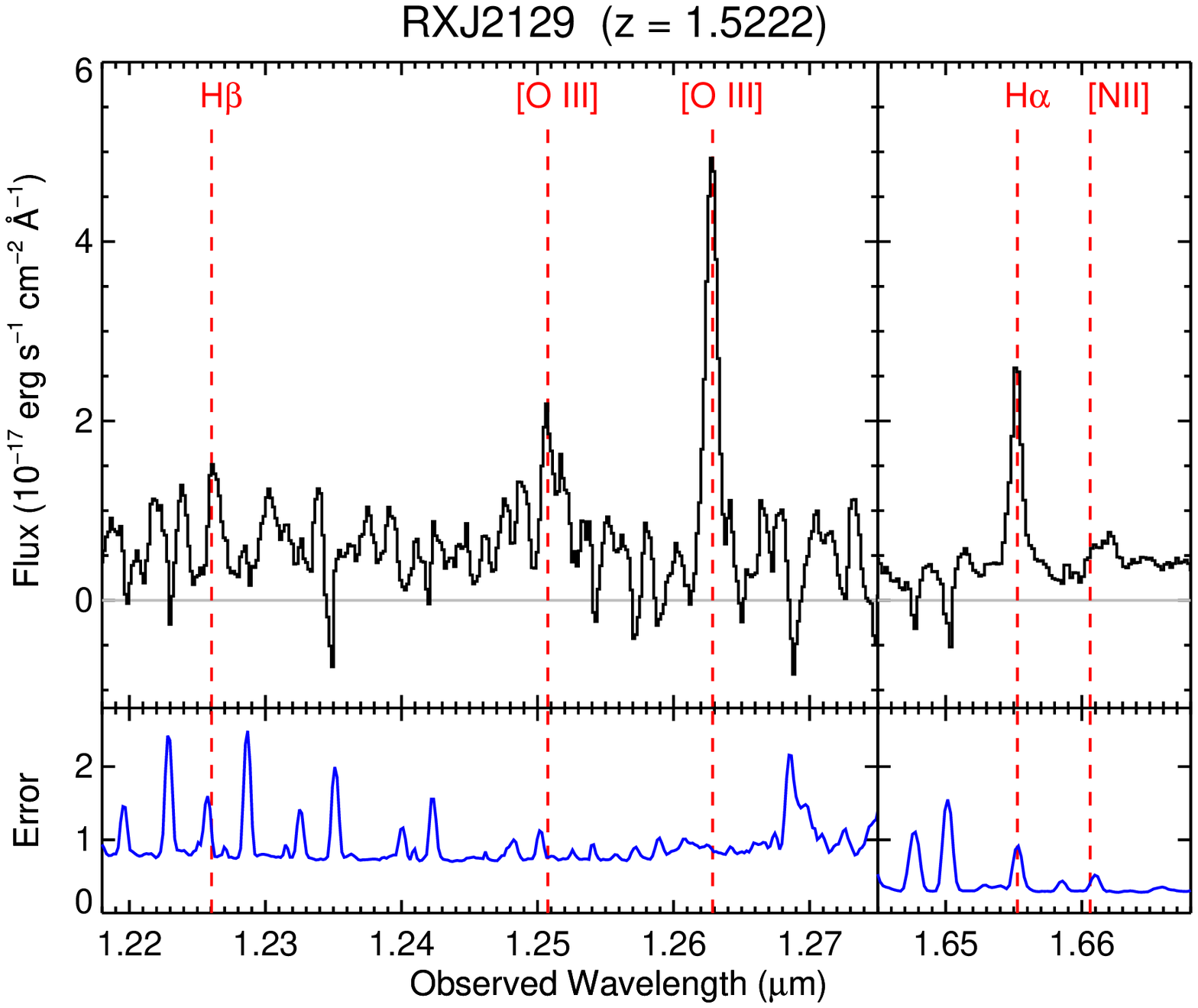}}
   \hspace*{0.02\textwidth}
 \raisebox{-0.5\height}{\includegraphics[width=0.32\textwidth]{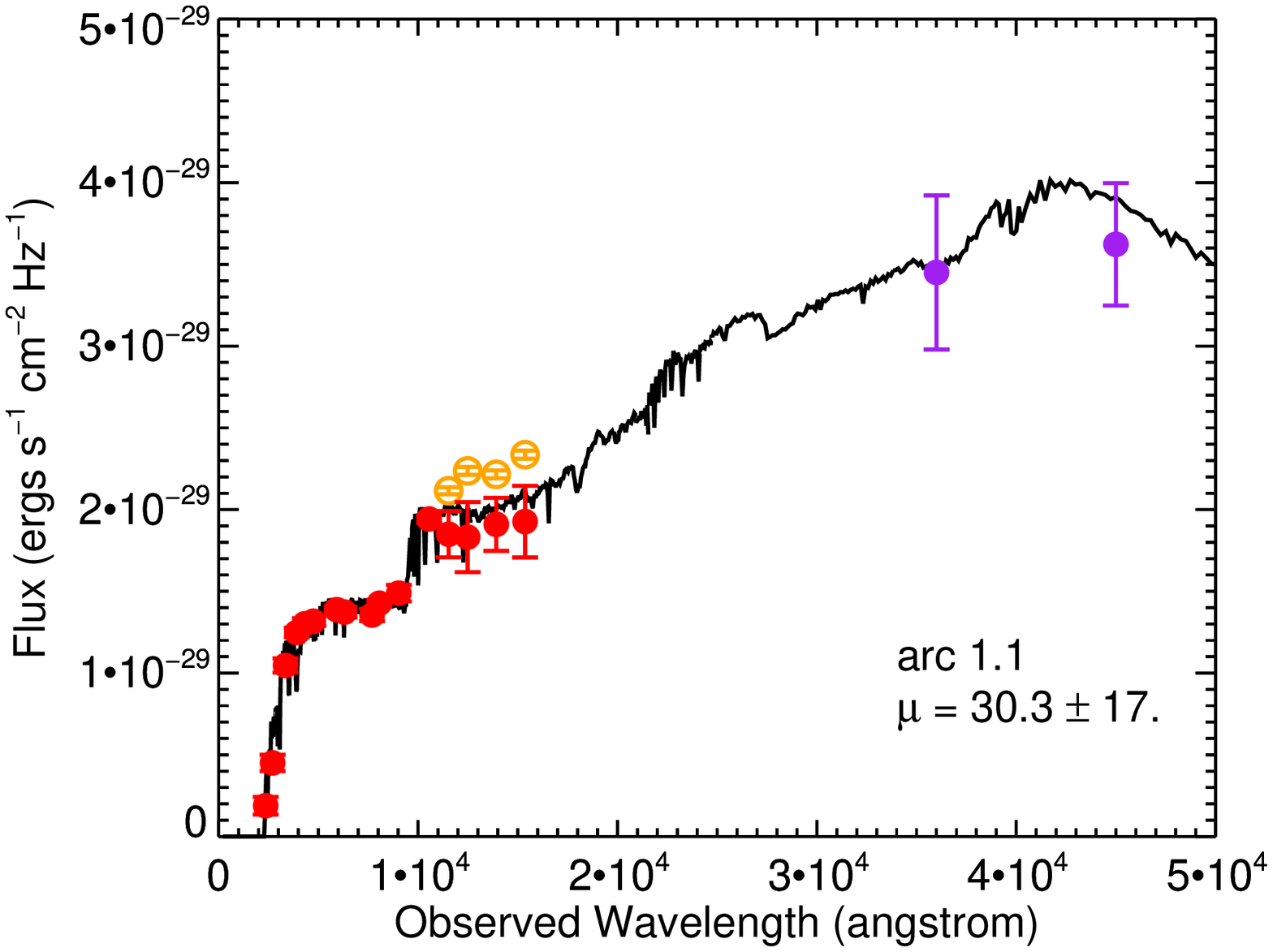}}
\end{minipage}

  \vspace*{0.02\textwidth}

\begin{minipage}{\textwidth}
   \centering
 \raisebox{-0.5\height}{\includegraphics[width=0.15\textwidth]{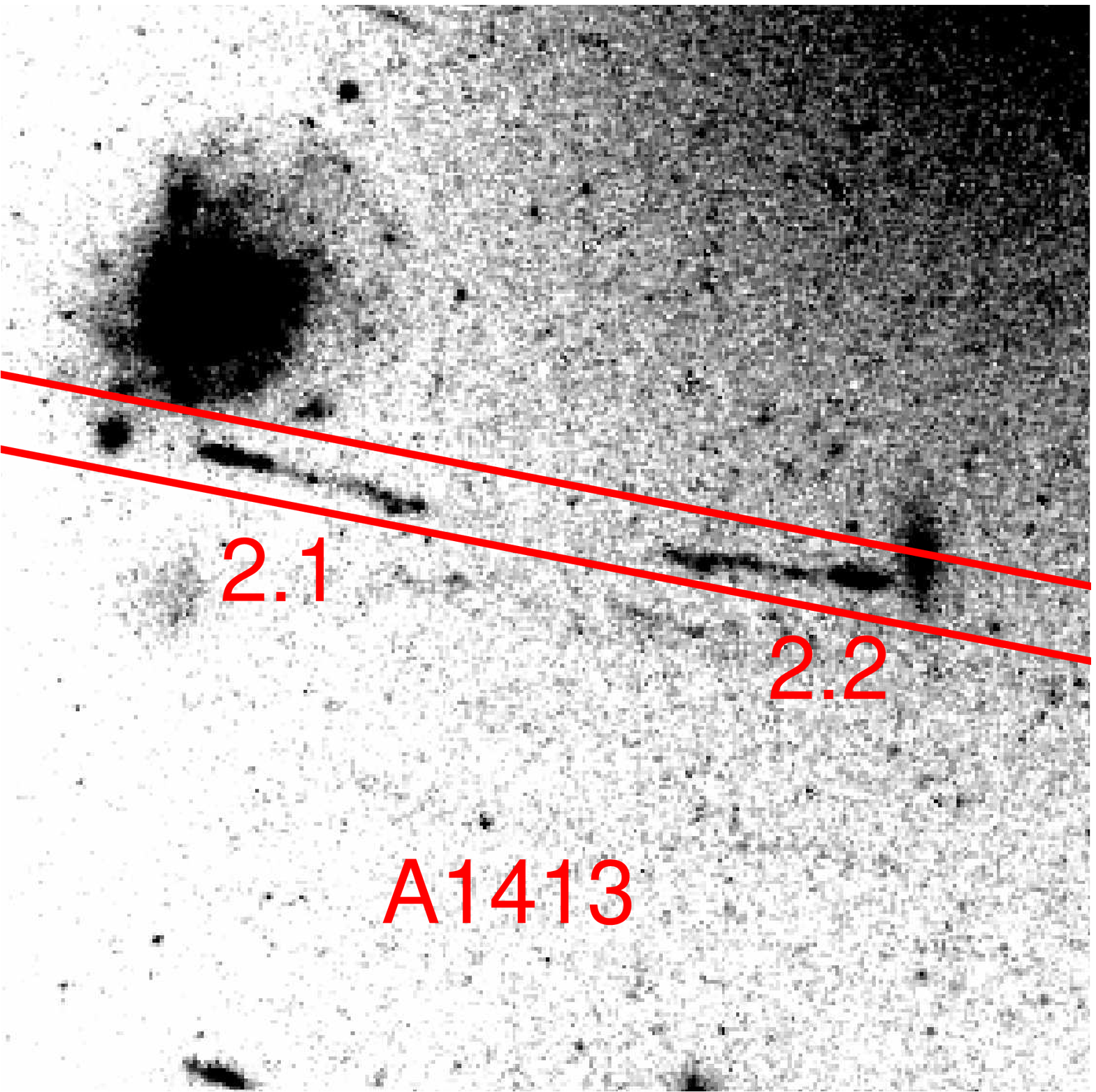}}
   \hspace*{0.02\textwidth}
 \raisebox{-0.5\height}{\includegraphics[width=0.32\textwidth]{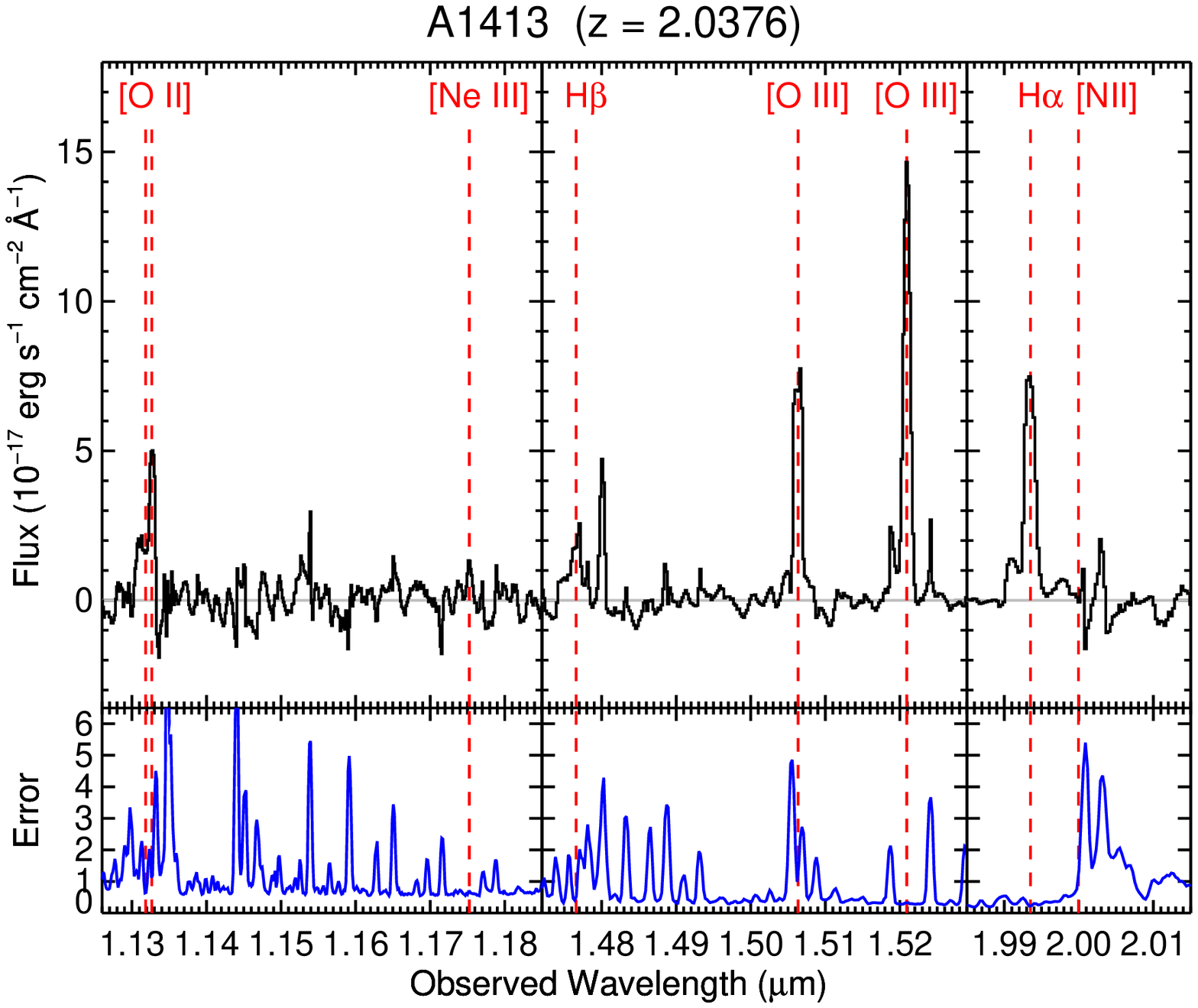}}
   \hspace*{0.02\textwidth}
 \raisebox{-0.5\height}{\includegraphics[width=0.32\textwidth]{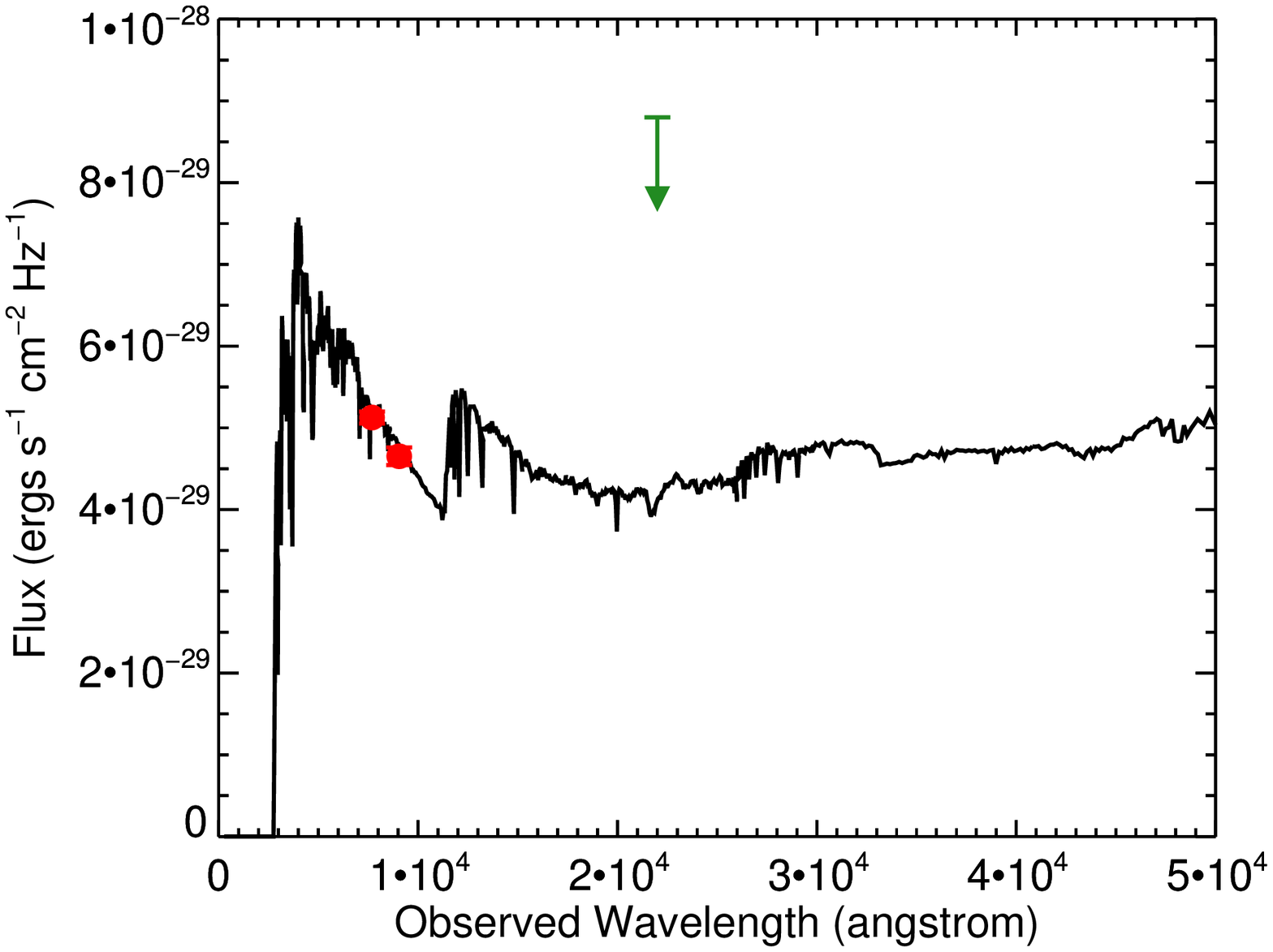}}
\end{minipage}

  \vspace*{0.02\textwidth}
  
\begin{minipage}{\textwidth}
   \centering
 \raisebox{-0.5\height}{\includegraphics[width=0.15\textwidth]{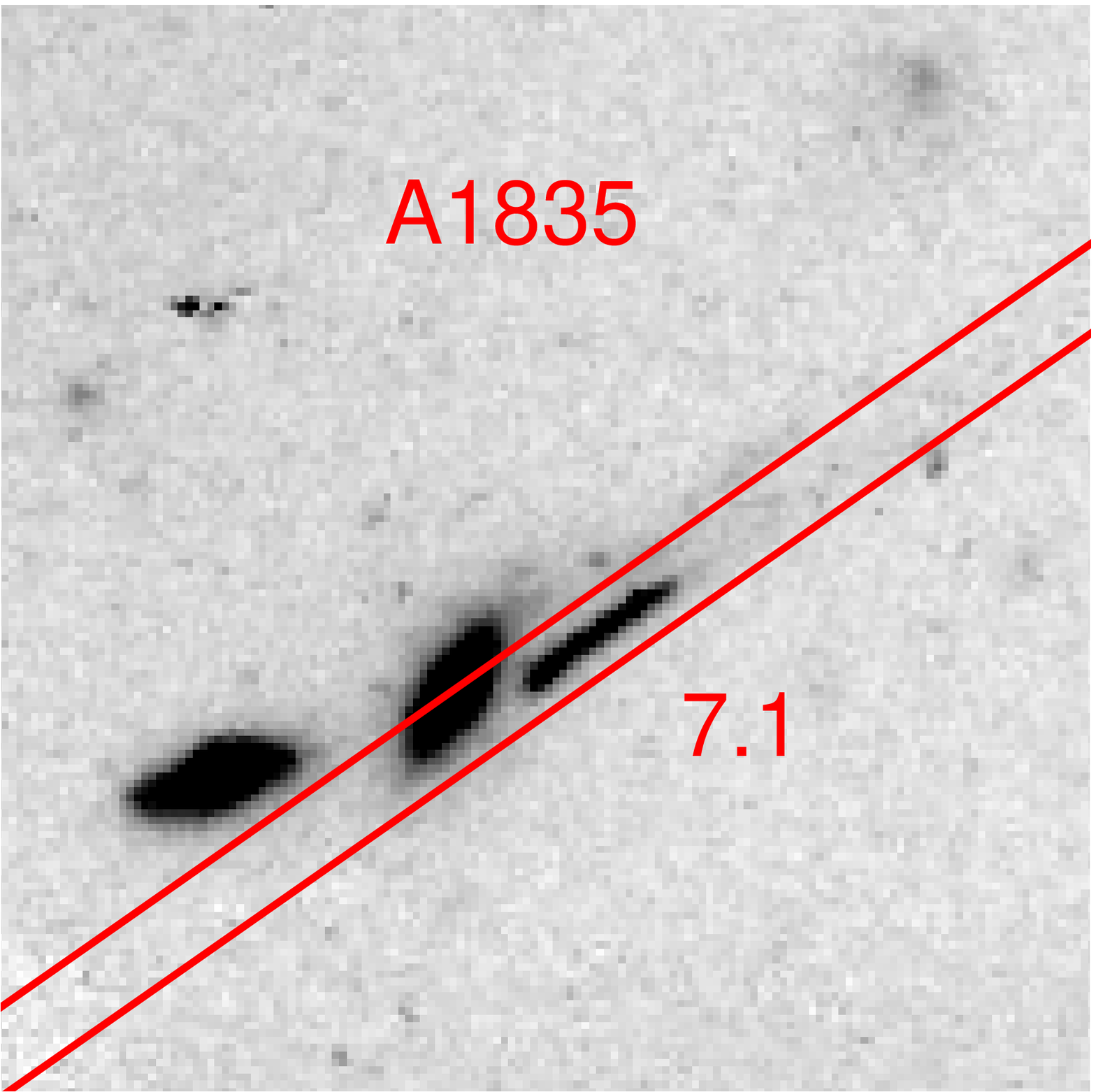}}
   \hspace*{0.02\textwidth}
 \raisebox{-0.5\height}{\includegraphics[width=0.32\textwidth]{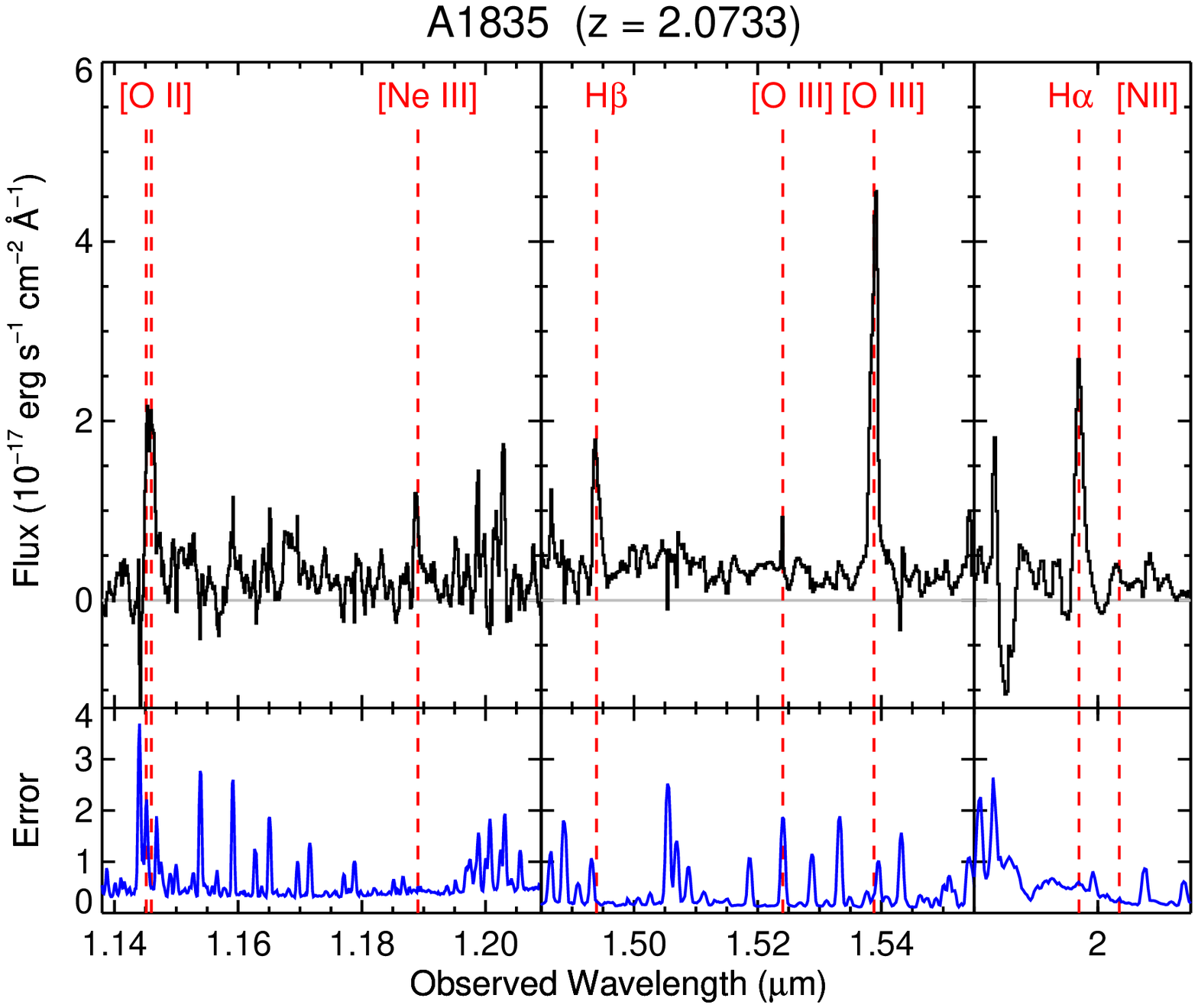}}
   \hspace*{0.02\textwidth}
 \raisebox{-0.5\height}{\includegraphics[width=0.32\textwidth]{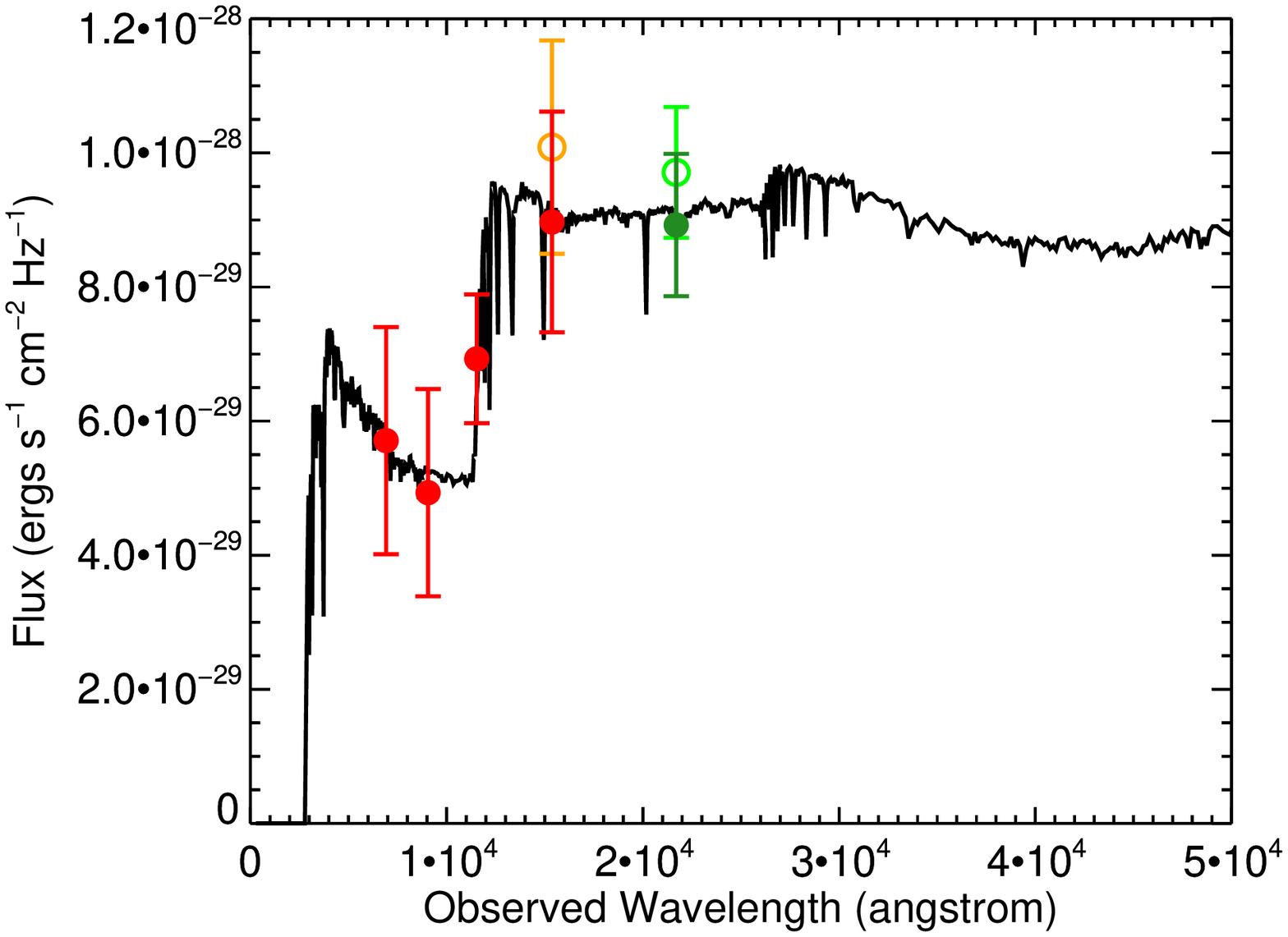}}
\end{minipage}

\caption{\HST\ image stamp, Triplespec spectrum, and SED fit for each gravitational arc. 
\emph{Left:} The \emph{Hubble Space Telescope} images (F702W or F775W) show the position of the Triplespec slit. Each square is 15 arcseconds on the side.
\emph{Center:} Each spectrum has been inverse-variance smoothed using a 5-pixel window, and the bottom panels show the 1-$\sigma$ error in the same units as the flux. When multiple images are present on the slit, the spectrum shown is the combined spectrum.
\emph{Right:} The photometry from (observed) UV to infrared is plotted as filled points, and the color corresponds to the type of data: red for \HST, green for ground-based near-infrared, and purple for \emph{Spitzer} IRAC images. The empty, lighter-colored points are photometric measurements not corrected for emission line flux (see Section \ref{sec:SED}), and the best-fit synthetic spectrum is shown as a solid line. For objects with two gravitational images, the SED is plotted only for the one that is less contaminated by foreground galaxies, and its magnification factor is shown.
Spectra and photometry are not corrected for the lensing magnification.
}
\label{fig:centerpiece}
\end{figure*}

\begin{figure*}[htbp]
    
\begin{minipage}{\textwidth}
   \centering
 \raisebox{-0.5\height}{\includegraphics[width=0.15\textwidth]{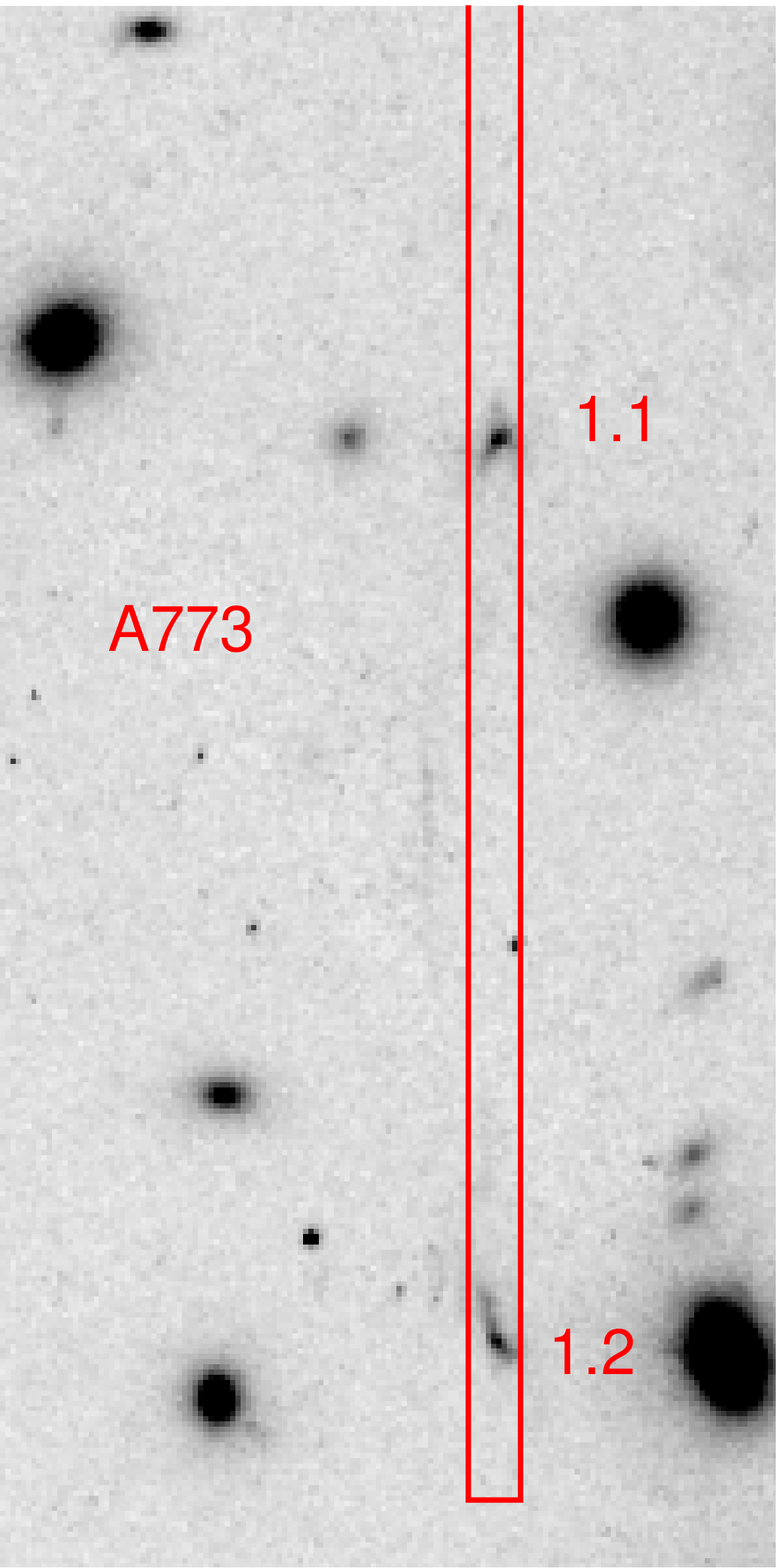}}
   \hspace*{0.02\textwidth}
 \raisebox{-0.5\height}{\includegraphics[width=0.32\textwidth]{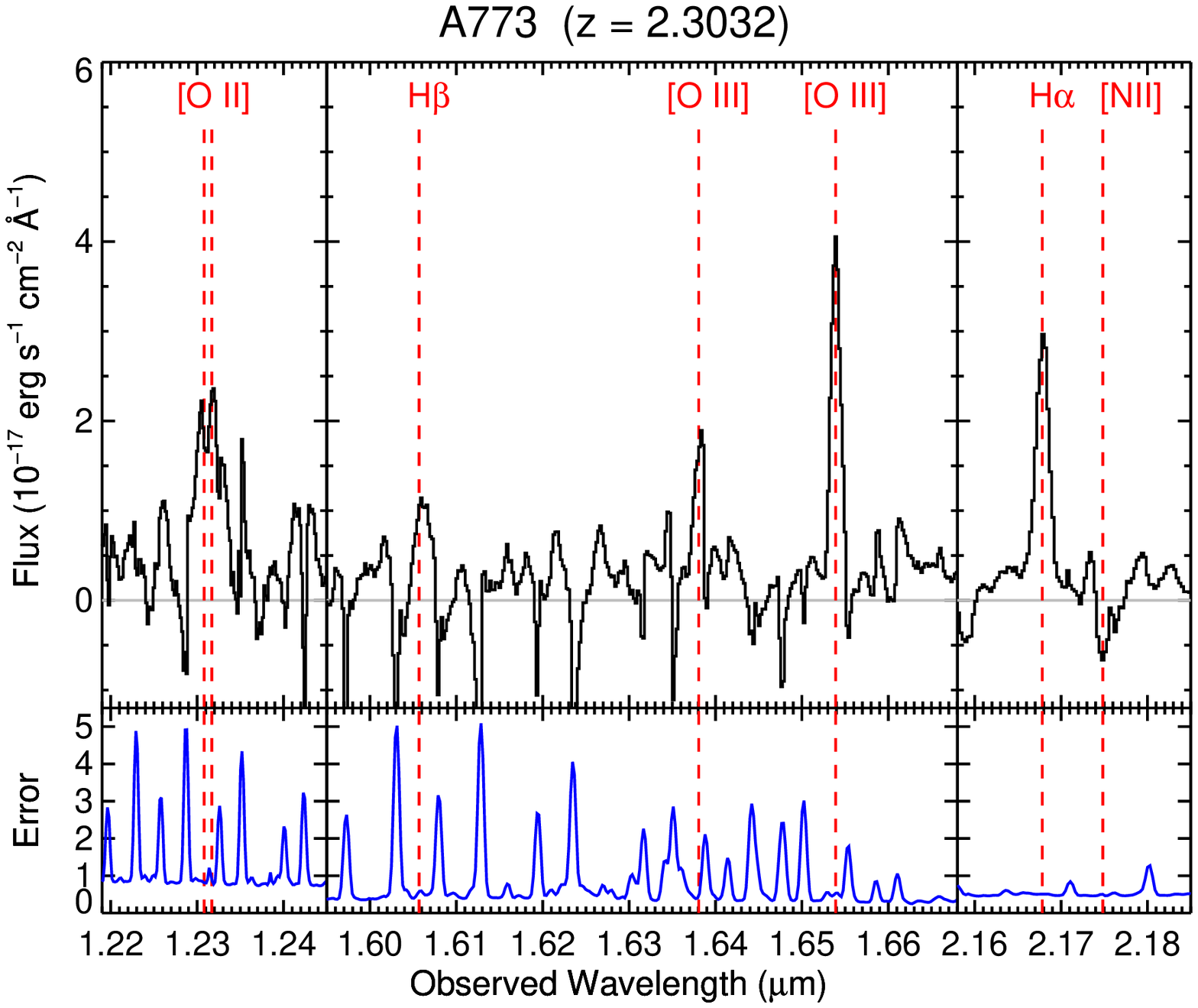}}
   \hspace*{0.02\textwidth}
 \raisebox{-0.5\height}{\includegraphics[width=0.32\textwidth]{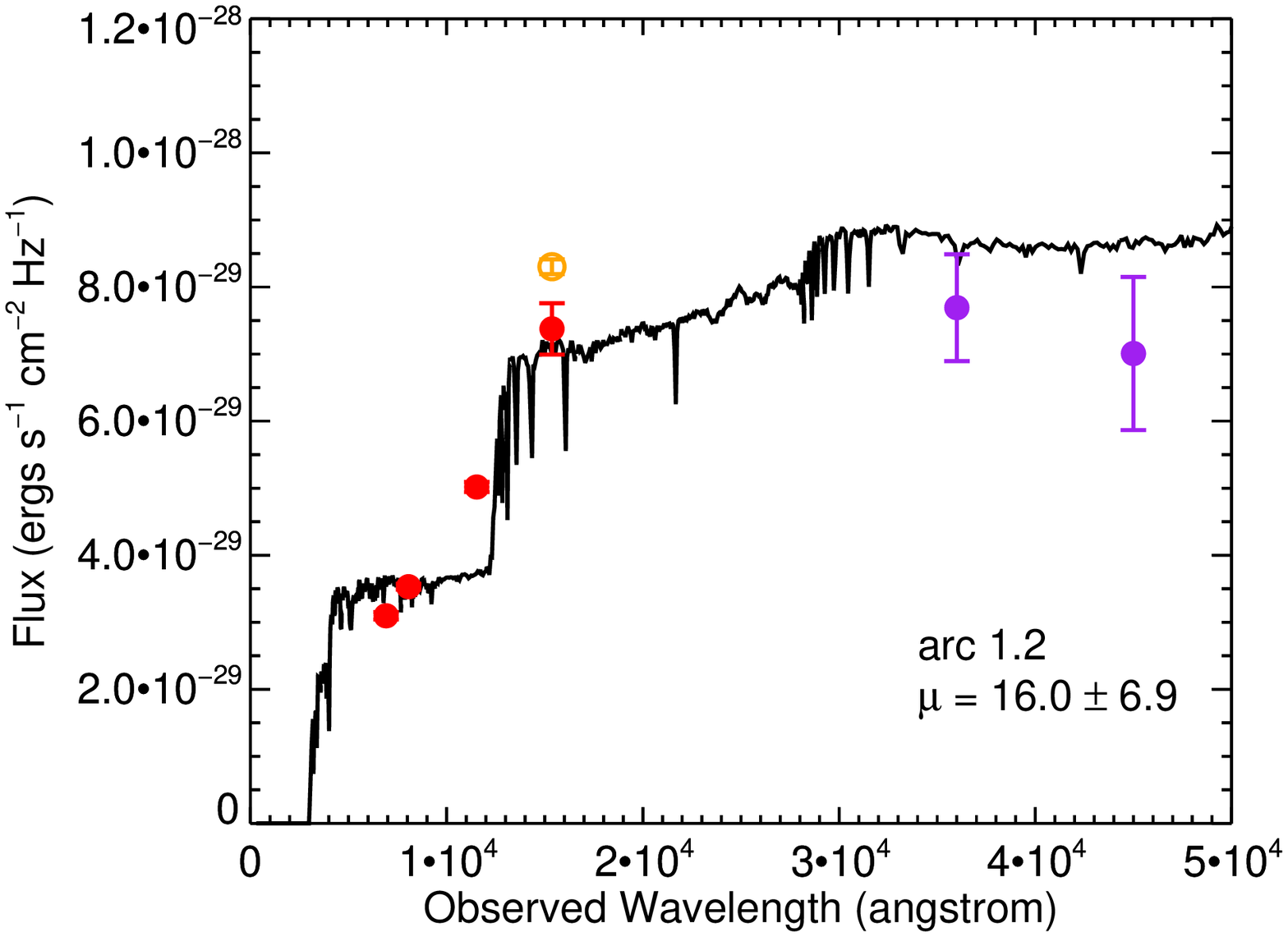}}
\end{minipage}

  \vspace*{0.02\textwidth}
  
\begin{minipage}{\textwidth}
   \centering
 \raisebox{-0.5\height}{\includegraphics[width=0.15\textwidth]{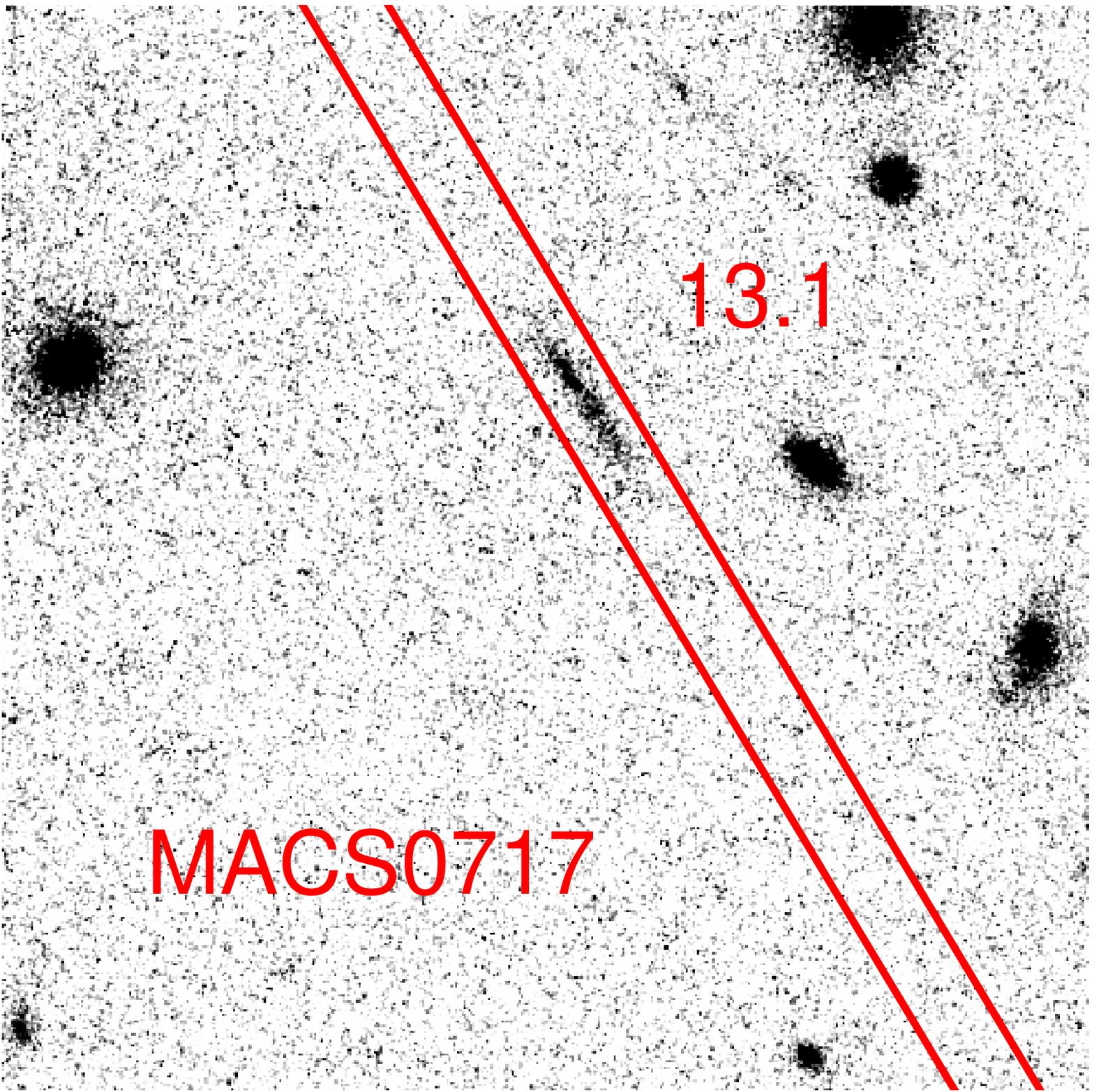}}
   \hspace*{0.02\textwidth}
 \raisebox{-0.5\height}{\includegraphics[width=0.32\textwidth]{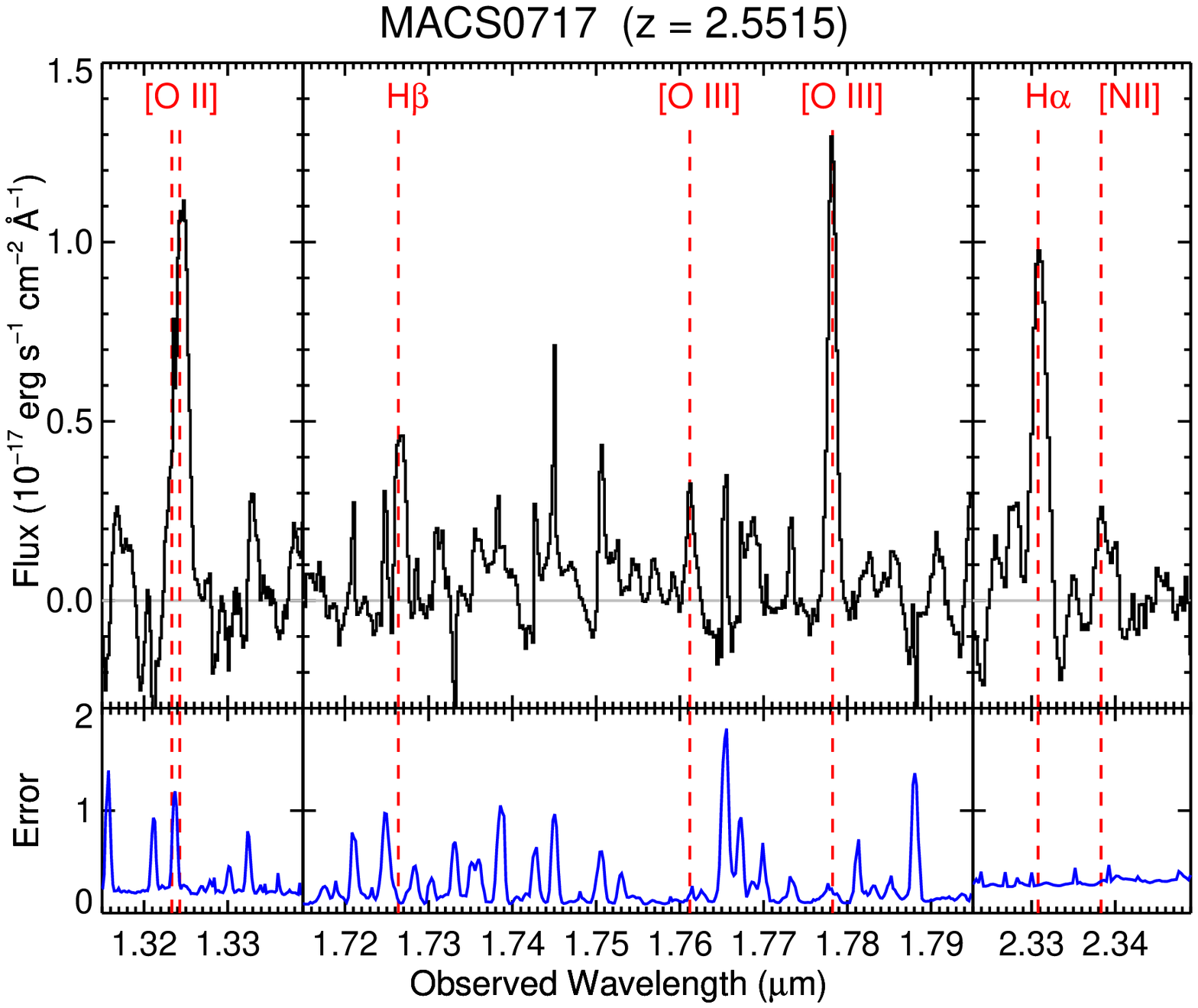}}
   \hspace*{0.02\textwidth}
 \raisebox{-0.5\height}{\includegraphics[width=0.32\textwidth]{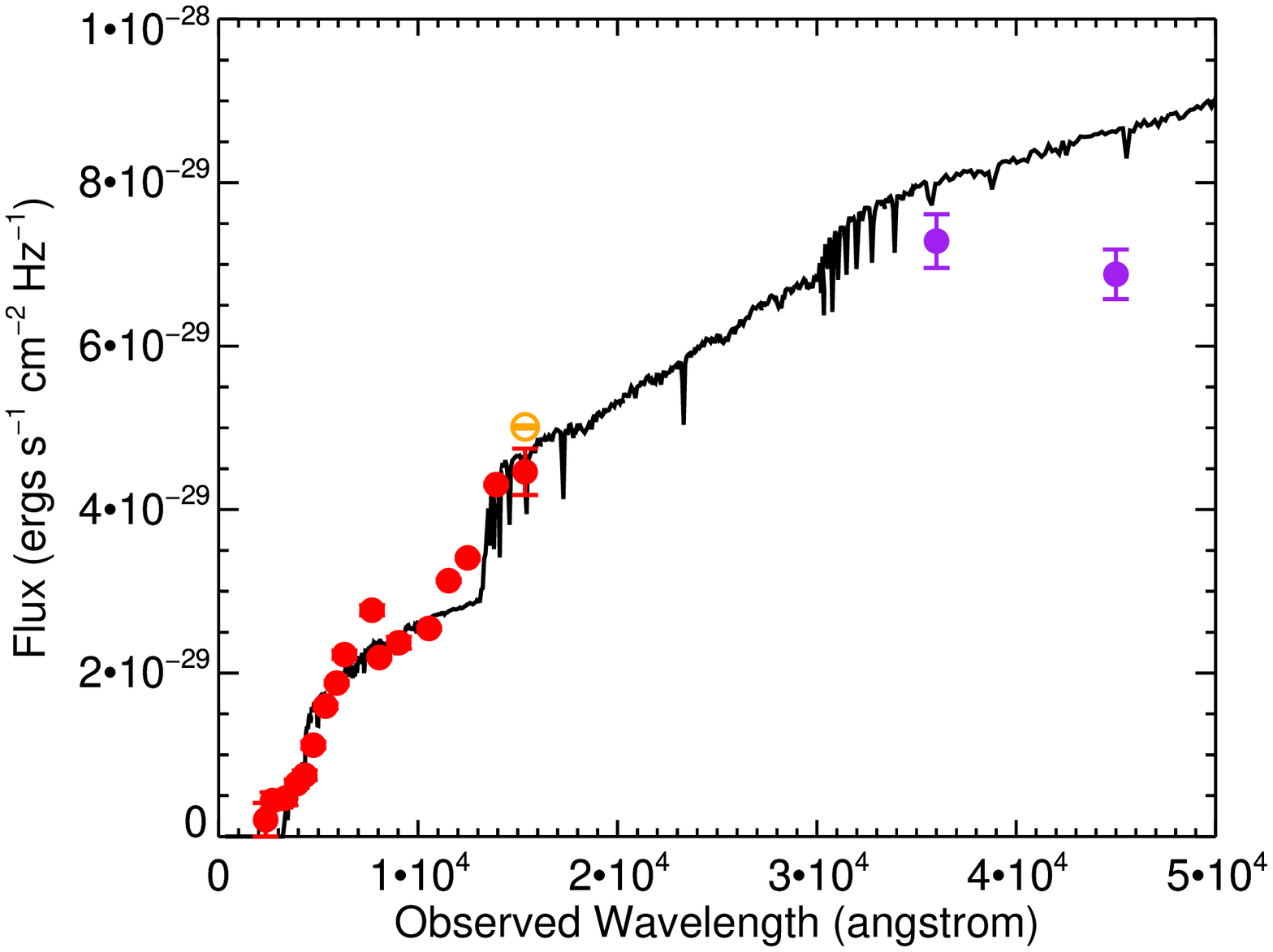}}
\end{minipage}

  \vspace*{0.02\textwidth}
  
\begin{minipage}{\textwidth}
   \centering
 \raisebox{-0.5\height}{\includegraphics[width=0.15\textwidth]{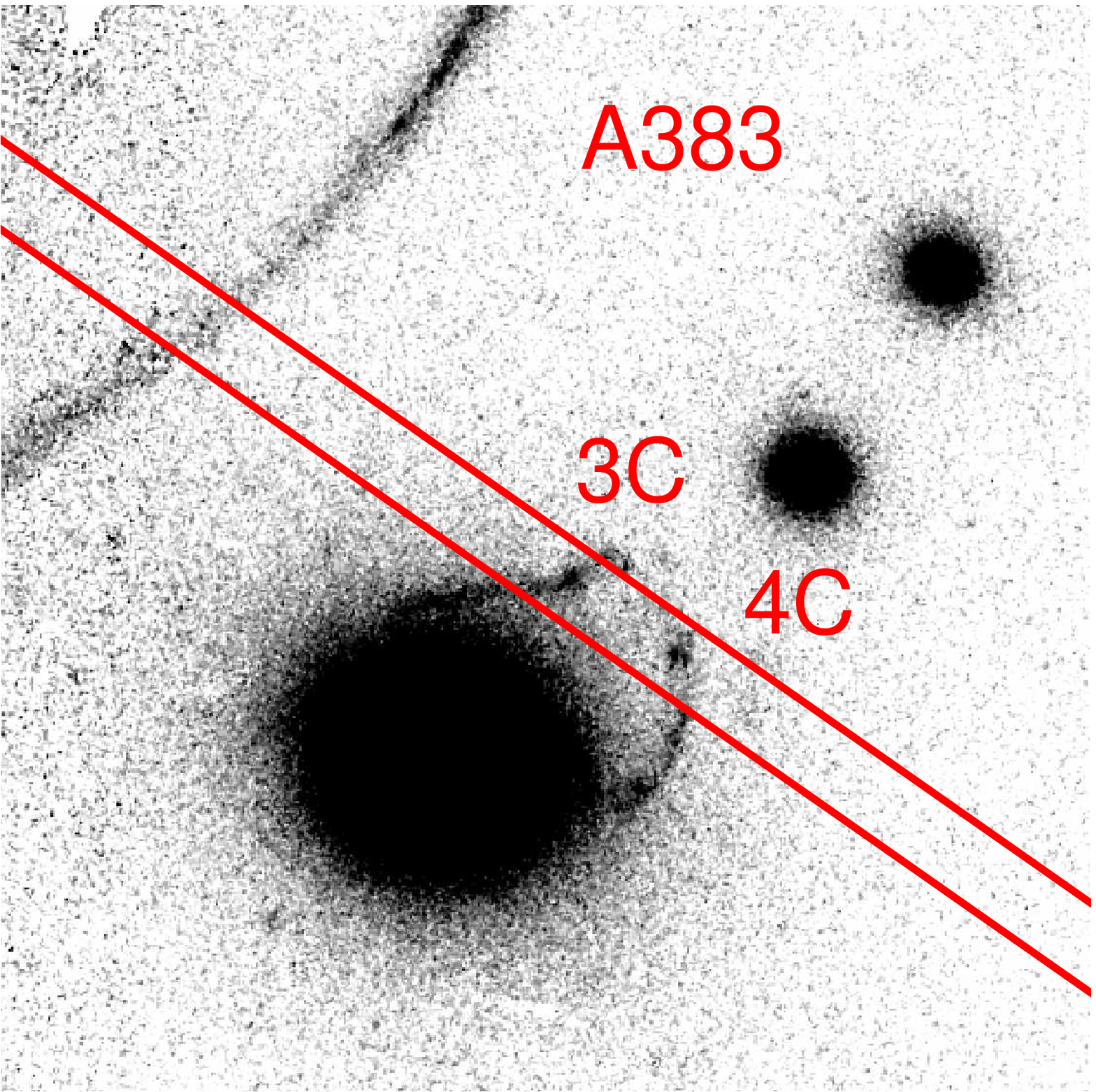}}
   \hspace*{0.02\textwidth}
 \raisebox{-0.5\height}{\includegraphics[width=0.32\textwidth]{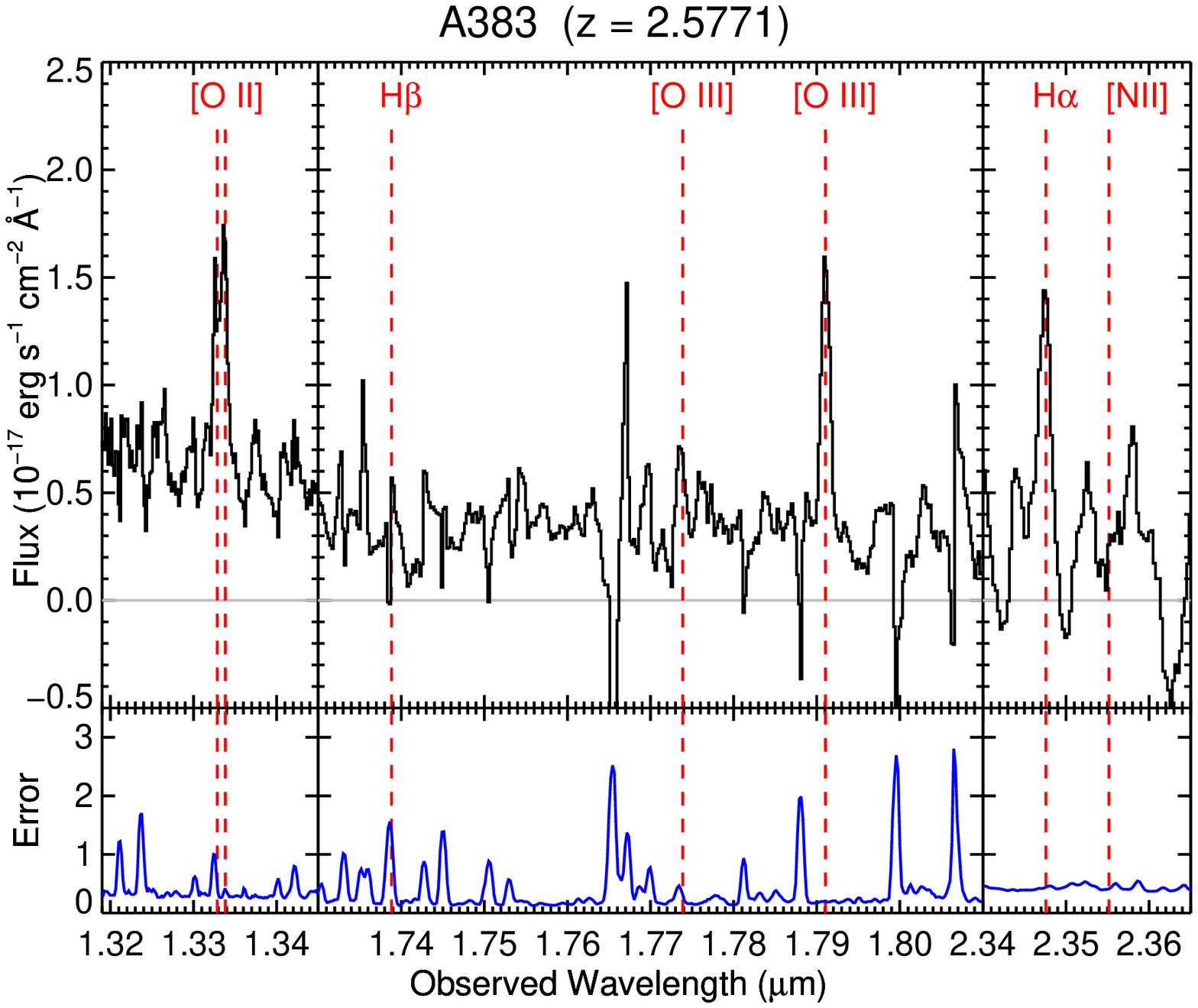}}
   \hspace*{0.02\textwidth}
 \raisebox{-0.5\height}{\includegraphics[width=0.32\textwidth]{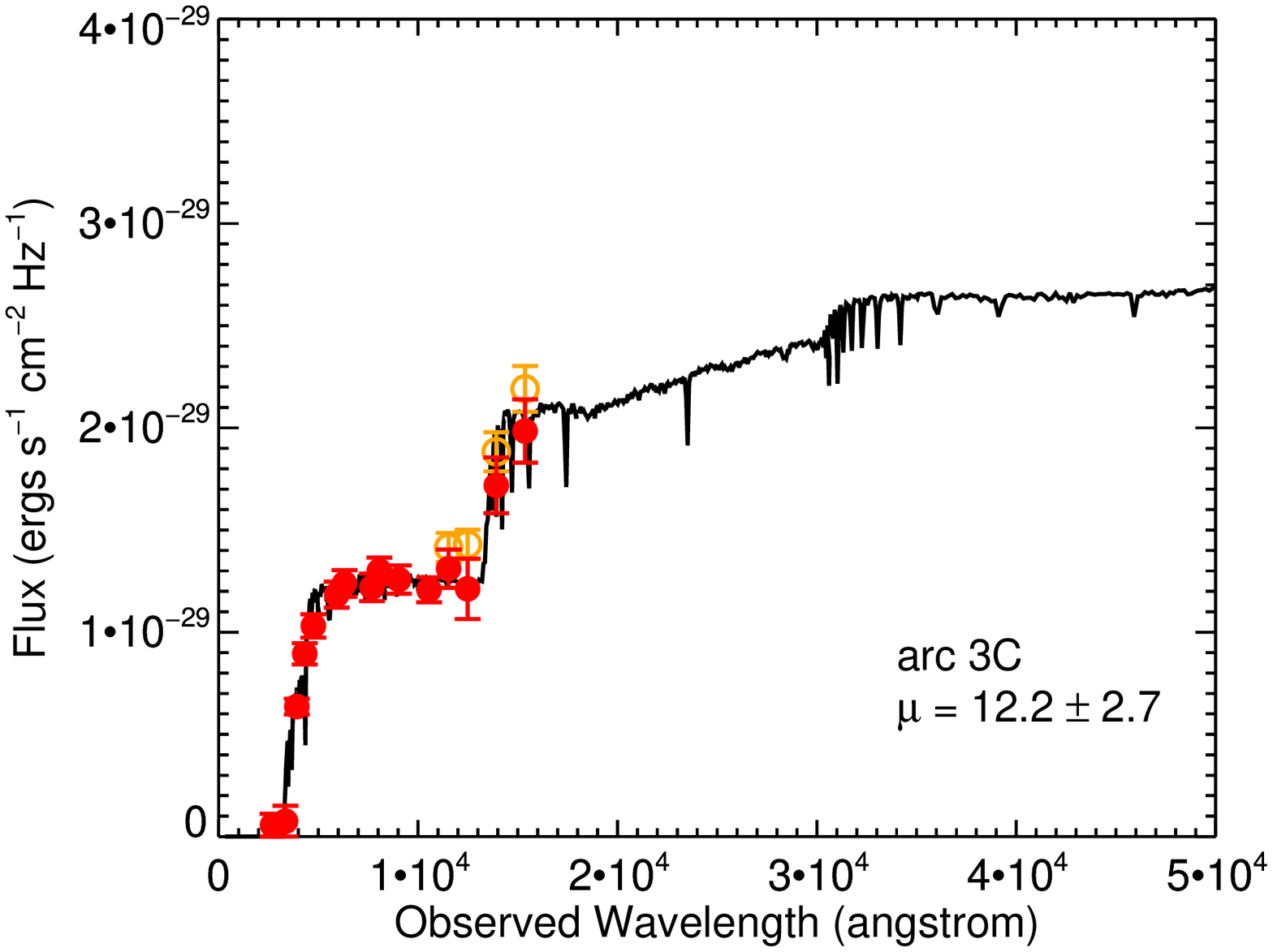}}
\end{minipage}

  \vspace*{0.02\textwidth}
  
\begin{minipage}{\textwidth}
   \centering
 \raisebox{-0.5\height}{\includegraphics[width=0.15\textwidth]{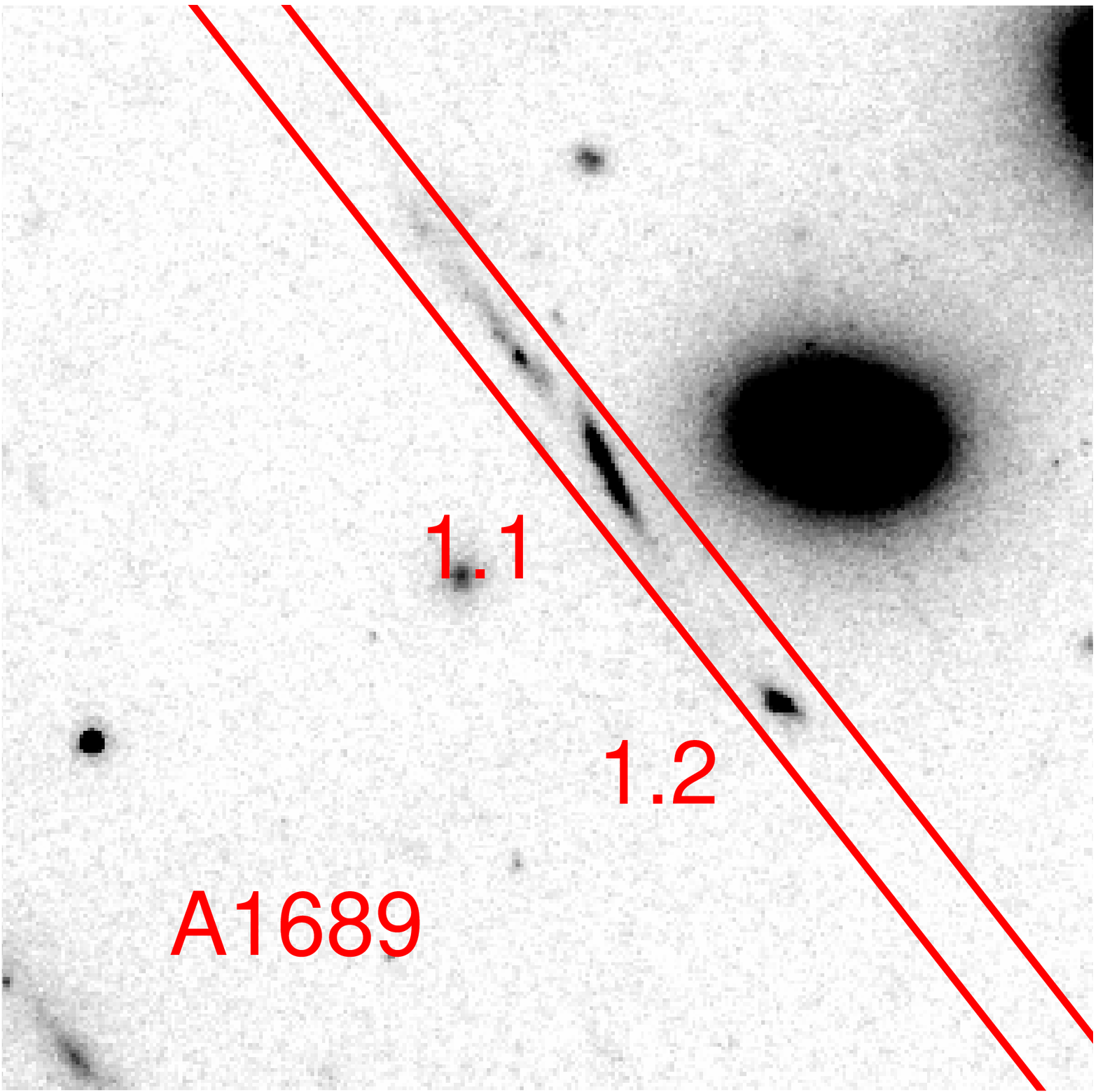}}
   \hspace*{0.02\textwidth}
 \raisebox{-0.5\height}{\includegraphics[width=0.32\textwidth]{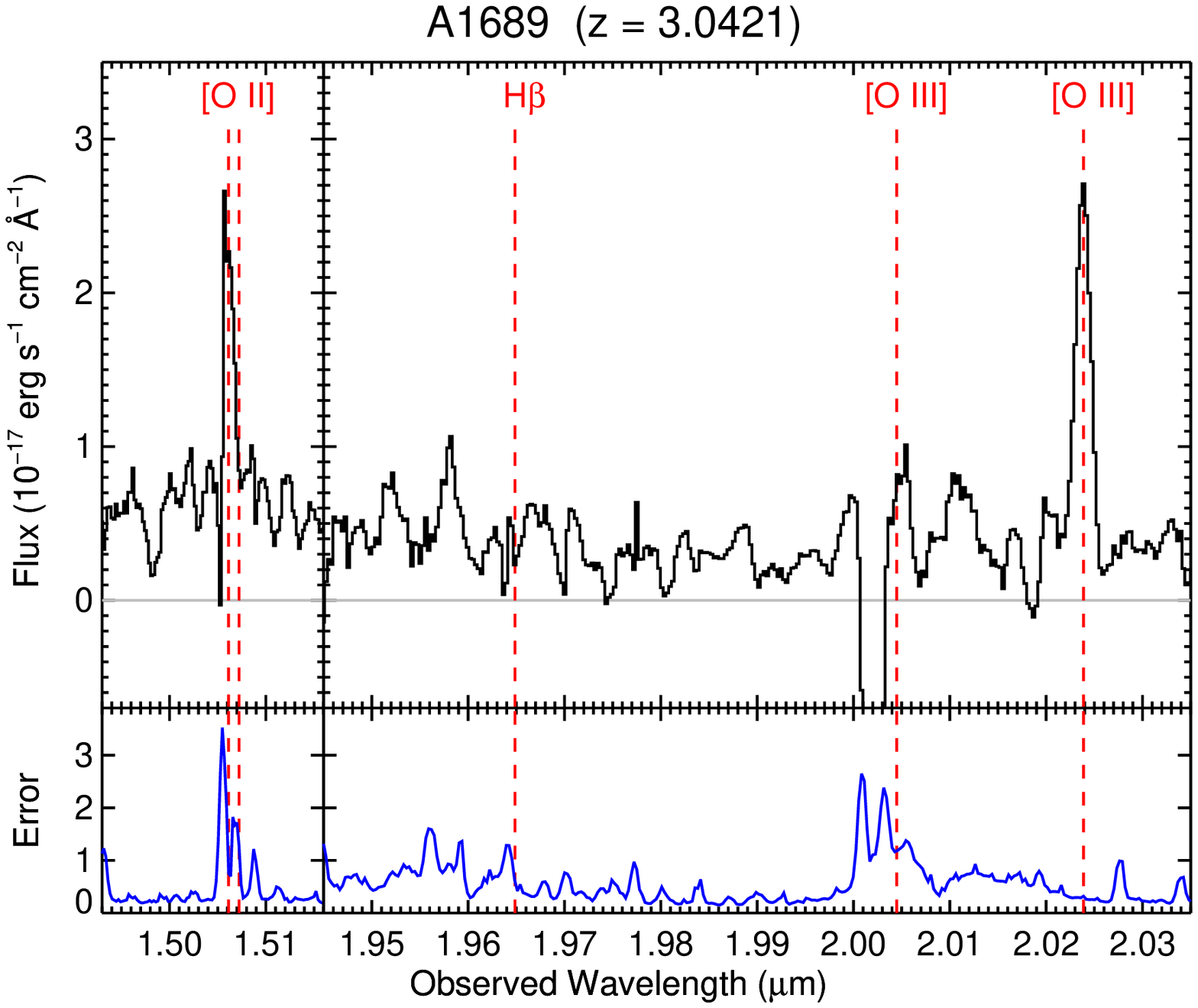}}
   \hspace*{0.02\textwidth}
 \raisebox{-0.5\height}{\includegraphics[width=0.32\textwidth]{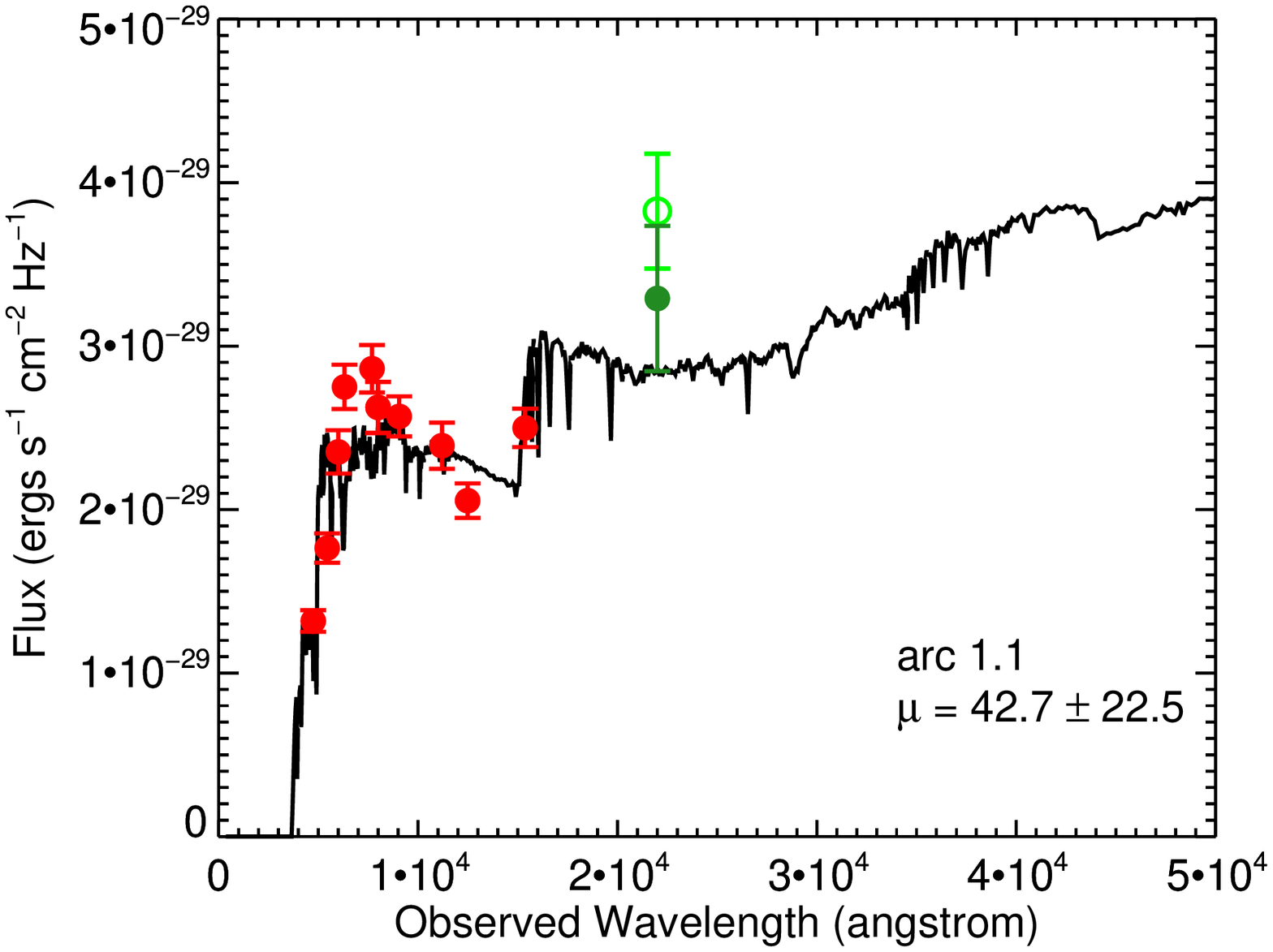}}
\end{minipage}
\addtocounter{figure}{-1}
\caption{Continued.}
\label{fig:centerpiece2}
\end{figure*}

\begin{figure*}[htbp]

\begin{minipage}{\textwidth}
   \centering
 \raisebox{-0.5\height}{\includegraphics[width=0.15\textwidth]{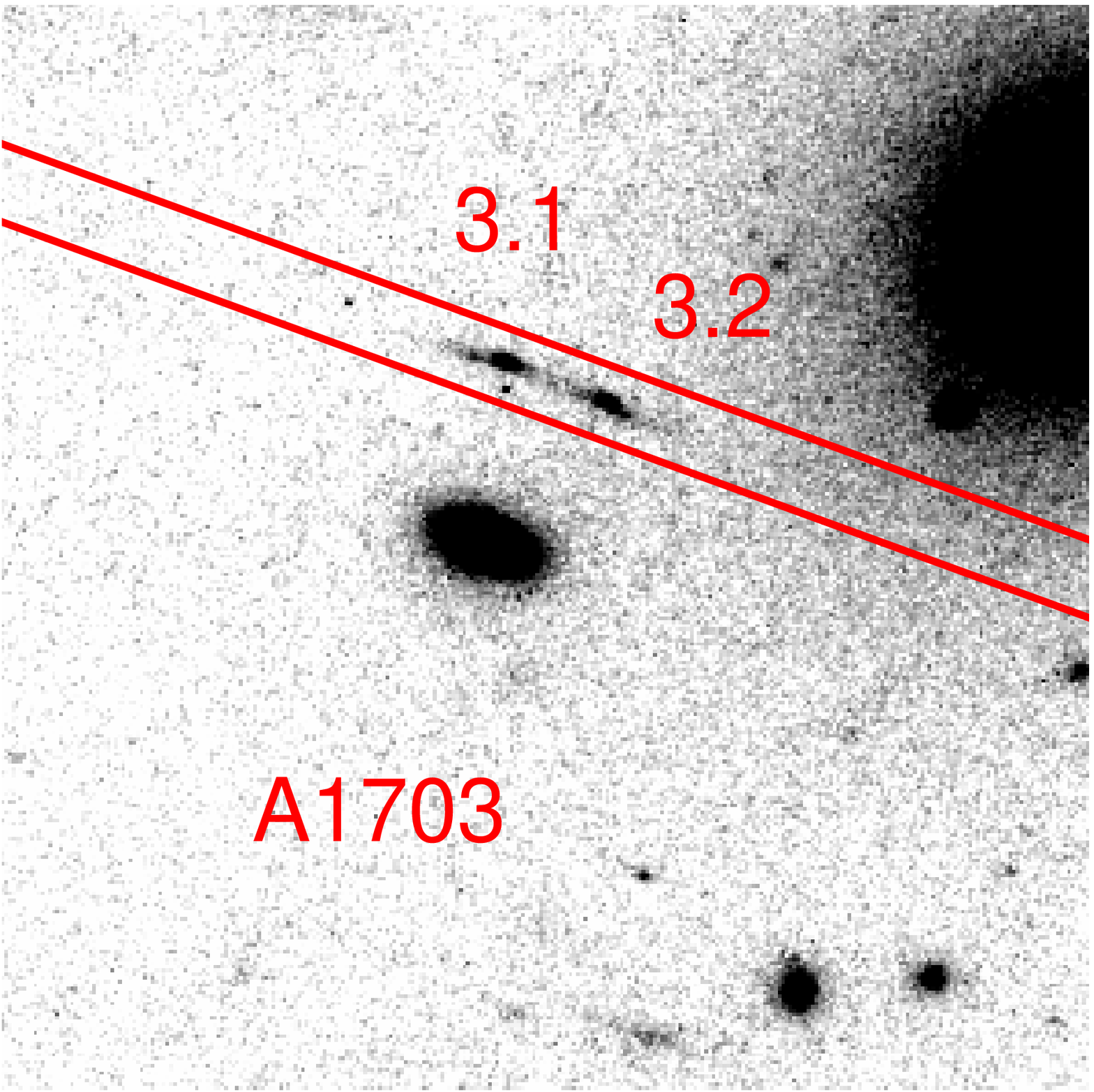}}
   \hspace*{0.02\textwidth}
 \raisebox{-0.5\height}{\includegraphics[width=0.32\textwidth]{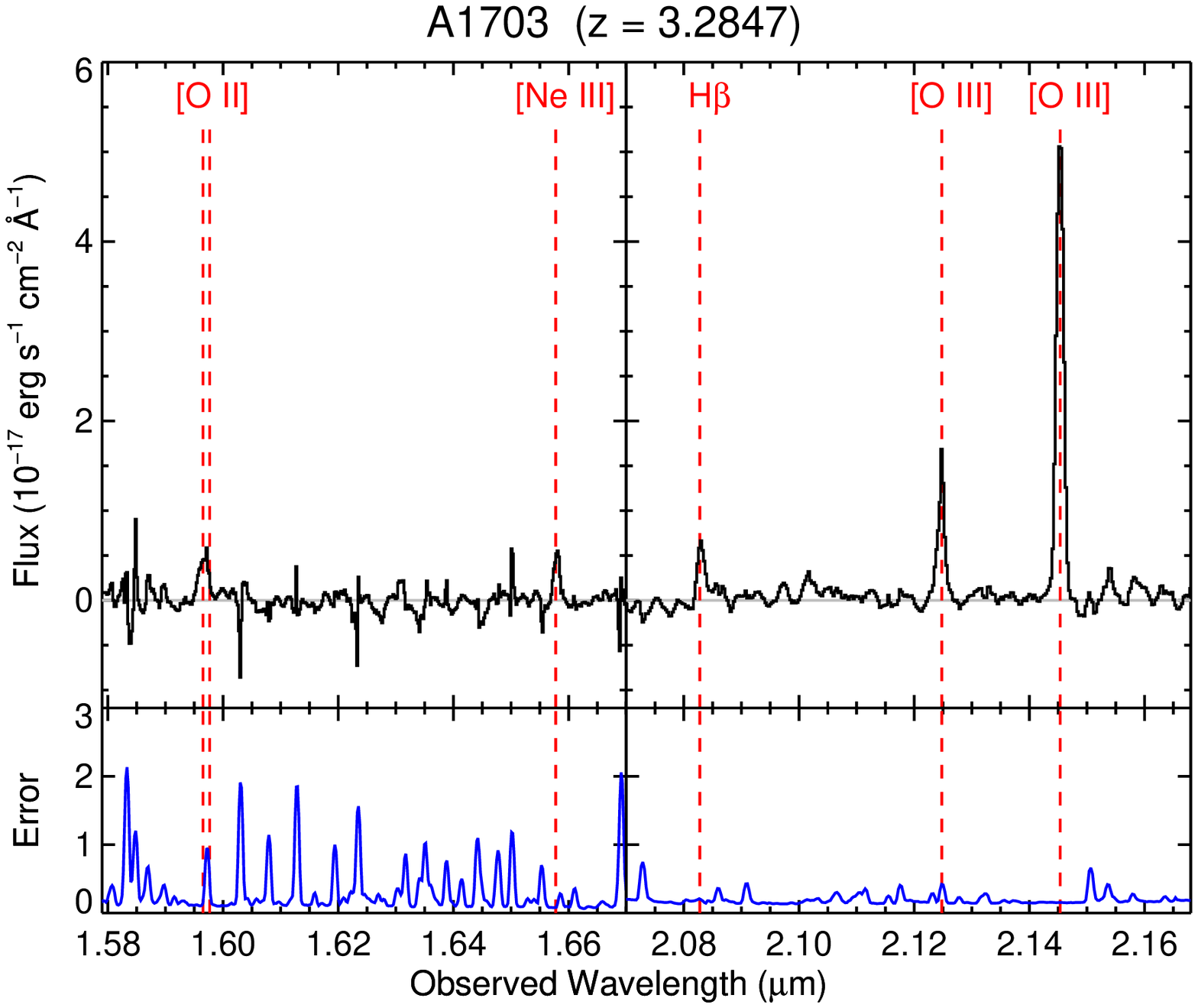}}
   \hspace*{0.02\textwidth}
 \raisebox{-0.5\height}{\includegraphics[width=0.32\textwidth]{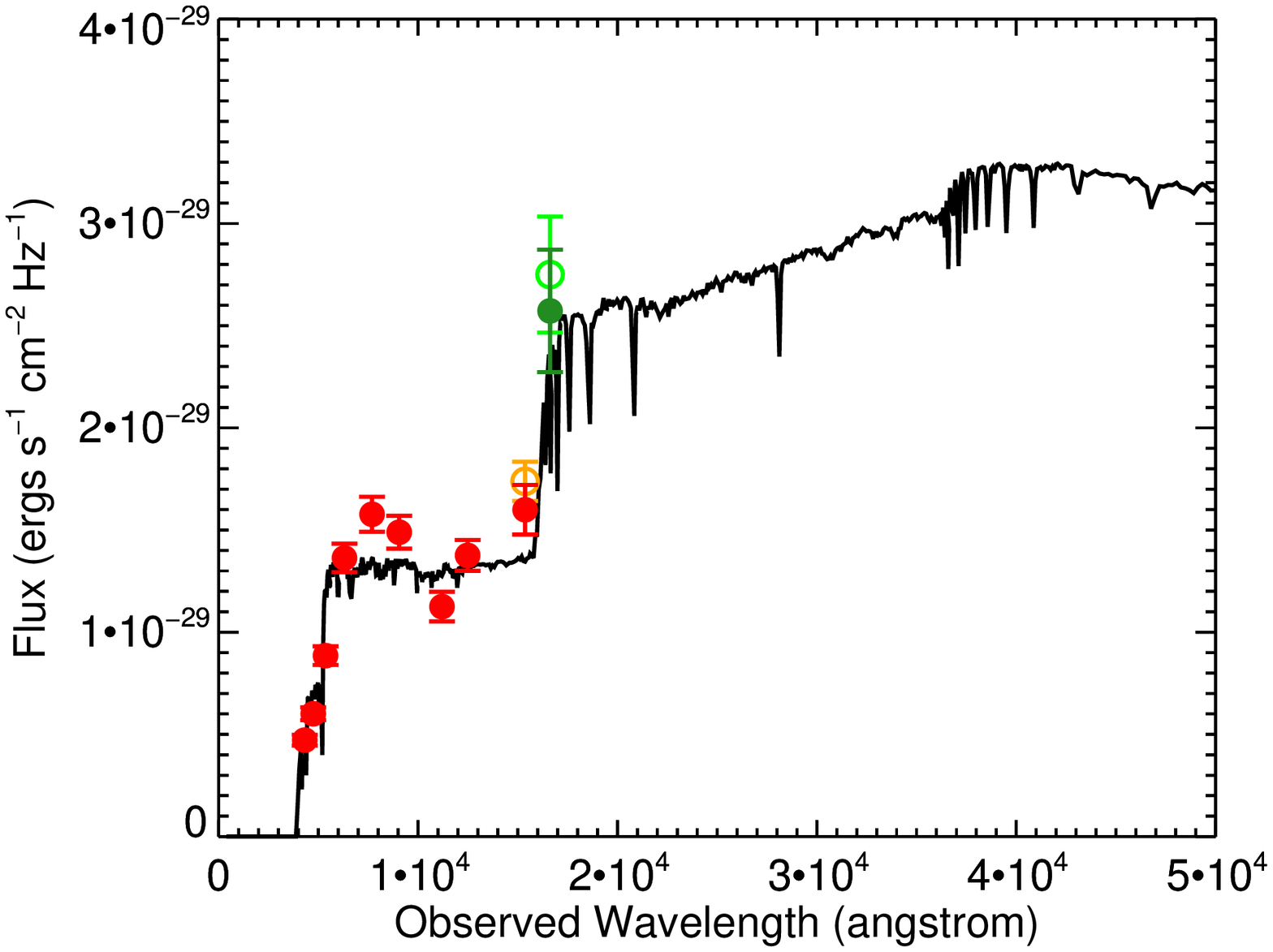}}
\end{minipage}

\addtocounter{figure}{-1}
\caption{Continued.}
\label{fig:centerpiece3}
\end{figure*}

The accuracy of the absolute flux calibration is limited by many factors. First, the observing conditions were not always photometric, and a good fraction of our observations were affected by the presence of thin clouds. This problem is mitigated by the fact that each target was observed on more than one night.

The second issue arises from the fact that every few minutes the pointing was changed according to the dithering pattern. As a result, the slit alignment is not identical in each frame, and the observed flux may depend on how the target is centered. We tested the significance of this effect using the standard star observations, where the target is bright enough to compare the flux in different frames. The discrepancy in the absolute flux between different frames is typically much less than 50\%. This represents an upper limit on the flux uncertainty, since this random effect is attenuated by averaging together multiple frames for each standard star. Also, the science observations were made on much longer timescales, and the guiding was overall very stable, as we could check from the guider images taken during the exposure.

Another possible source of uncertainty is the variable seeing. However, the difference in seeing (which was almost always larger than the slit width) between the science target and the standard star observations has a much smaller effect than the slit misalignment.

For each object we separately flux-calibrated the spectra from different nights using the appropriate standard stars. This reduces the effect of seeing variation and cloud attenuation. We then measured the flux of the brightest line. Since slit misalignment and clouds tend to attenuate the line emission, we scale the spectra from different nights to match the one with the brightest line. For each object we have at least some observations with clear conditions, so that the flux uncertainty caused by cloud cover is negligible compared to the $50\%$ uncertainty measured from the standard star misalignment. This represents therefore a conservative estimate of the overall flux calibration error. The corresponding contribution to the uncertainty in the SFR is 0.22 dex.

The calibrated spectra together with their 1-$\sigma$ error are shown in Fig. \ref{fig:centerpiece}. The emission line \OIIIb\ is well-detected in each spectrum; other observed emission lines are \Halpha, \Hbeta, \OIIIa, and \OII. The fainter lines \NII\ and \NeIII\ are detected only in a few cases. The continuum emission from the arcs is never detected, but the residuals from sky subtraction or the emission from foreground galaxies can cause the observed continuum to be different from zero.

\subsection{Imaging}

We now discuss the imaging data for our sources which will provide the essential ingredients for measuring the stellar masses and other physical properties. To accurately derive the stellar mass it is necessary to sample the spectral energy distribution (SED) redward of the rest-frame Balmer break. For $z<2$ objects optical imaging is sufficient, but at higher redshift infrared data are needed.

\subsubsection{Archival Data}

Since we chose well-studied galaxy clusters, space-based images from \emph{Hubble Space Telescope} (\HST) and \emph{Spitzer} observations are available for each target in our sample.

For four of the objects (A383, MACS0717, RXJ2129, and A611) we used publicly available data from the Cluster Lensing and Supernova Survey with Hubble \citep[CLASH,][]{postman12}. This allowed us to measure the photometry for each arc in about 16 bands from UV to near-infrared. For the remaining targets we used archival \HST\ images; each source has imaging available in at least four bands except A1413, for which only two bands are available.

All of the selected targets have publicly available \emph{Spitzer} IRAC observations, but due to the faintness of the gravitational arcs not all of them are detected. The arcs with useful IRAC imaging are A611, RXJ2129, A773, and MACS0717. We used channel 1 (3.6 \microm) and channel 2 (4.5 \microm) observations from \emph{Spitzer} program 60034 (PI: E. Egami) for all of the arcs, and data from program 83 (PI: G. Rieke) for A773.

Additionally, some ground-based near-infrared data have been used for a few arcs: VLT ISAAC $K_S$-band for A1835 \citep{richard06} and A1689 (J. Richard et al., in preparation), and Subaru MOIRCS $H$-band for A1703 \citep{richard09}.

\subsubsection{Palomar Observations}

Using ground and space-based archival data we can probe the spectral region redward of the Balmer break for each arc, with the exclusion of A1413. Since this type of photometry is crucial for measuring the stellar mass, we took $K_S$-band imaging for A1413 using WIRC on the Palomar 200-inch telescope. We used a 9-point dithering pattern of two-minute frames for a total exposure time of 198 minutes.


\section{Properties of Stellar Populations}

\subsection{Photometry}

Since the arcs experience a large gravitational magnification, they tend to lie at short projected distances from foreground galaxies. It may then be necessary to subtract the light of the cluster galaxies in order to reliably measure the arc photometry. In these cases we used Galfit \citep{peng02}, which allowed us to simultaneously fit many galaxies while taking into account the Point Spread Function (PSF) of the instrument, which was measured from bright, isolated stars in the field. This is particularly important when working with \emph{Spitzer} IRAC images, which present a very large and asymmetric PSF.

Sometimes the gravitational arc emission is contaminated by the cluster bright central galaxy (BCG) luminous halo, and as a result the local background is hard to model. In these cases we fit the BCG using the Multi-Gaussian Expansion algorithm developed by \citet{cappellari02}. This method consists of fitting the surface brightness of a galaxy with a series of two-dimensional Gaussian functions, and is very effective for bright, extended galaxies.

After the subtraction of the foreground galaxies, the arc photometry is measured using polygonal apertures. The major source of uncertainty is generally the modeling of foreground galaxies and, for some IRAC images of crowded fields, confusion. The relative error in the flux is ${\sim} 30 \%$ in the worst cases, but typically much less than that. All the photometric measurements are corrected for galactic extinction according to the map of \citet{schlegel98}.

The photometric points are plotted for each arc in Figure \ref{fig:centerpiece} with a different color for each set of observations: red for \emph{Hubble Space Telescope}, green for ground-based near-infrared, and purple for \emph{Spitzer} data. When two gravitational images corresponding to the same source are available, the photometric analysis has been carried out independently on the different images, and the one less affected by foreground contamination was selected for the SED fitting. In these cases the name and the magnification factor of the chosen image are reported in Figure \ref{fig:centerpiece}.

The WIRC $K_S$-band data for A1413 are not deep enough to detect the faint gravitational arc, because the contamination from the foreground BCG galaxy's halo is very strong. In this case we can only derive an upper limit on the flux, and we show it in Figure \ref{fig:centerpiece}.

\subsection{SED Fitting}
\label{sec:SED}

For each target we fit the photometry from the observed UV to infrared using the stellar population models of \citet{bruzual03} in order to measure the stellar mass and other physical properties of the galaxies. We performed the fit using the chi-square minimization code FAST \citep{kriek09} assuming the \citet{calzetti00} dust extinction law and a \citet{chabrier03} initial mass function.

Since the emission lines detected in the Triplespec spectra are extremely bright compared to the continuum, we subtracted the measured line fluxes (see Section \ref{sec:spectra}) from the appropriate photometric bands. The errors in the absolute flux calibration (see Section \ref{sec:spectroscopy}) are propagated through the corrected photometry. For targets with multiple images on the slit the correction is calculated taking the flux from the combined spectrum, which has the advantage of a better signal to noise ratio, and then appropriately scaling it using the \HST\ photometry. The contribution of the emission lines can be as high as $40\%$ for a single band, but the effect on the stellar mass estimate is generally small: the average correction to the stellar mass is 0.10 dex. The only exception is A611, for which a combination of strong emission lines and small uncertainty on photometry causes the stellar mass to change by more than 2 sigma when applying the emission line correction.

Various stellar population parameters and degeneracies are involved in the process of SED fitting. One of the most important assumptions, and one that strongly affects the best-fit current SFR, is the star formation history. The widely used exponentially declining star formation history, or $\tau$-model, may not be an appropriate choice since these galaxies are young and have a large gas reservoir. Sometimes an inverted $\tau$-model is used for star-forming galaxies at high redshift. Both of these models require strong assumptions on the current state of the galaxy, i.e.\ that its star formation is currently at its minimum, or maximum, respectively. Also, they both introduce the free parameter $\tau$ which is usually not well-constrained by the data. For exponentially declining star formation histories, models with $\tau < 300$ Myr can give a formally acceptable fit to the data but usually fail in reproducing the SFR derived using other indicators \citep{wuyts11}. Since star-forming galaxies at high redshift are young, large values of $\tau$ imply a nearly flat star formation history. For these reasons we make the simplifying assumption of a constant star formation history. \citet{shapley05SED} consider a sample of star forming galaxies at $z \sim 2$ and show that the agreement between stellar masses derived assuming $\tau$-models or constant star formation history is very good, with no systematic offset and negligible dispersion. They also conclude that the choice of a particular SFH does not affect the uncertainty in the stellar mass measurement.

As shown by \citet{wuyts12SED}, the SED fit of low-mass star-forming galaxies tends generally to favor extremely young ages. Since a galaxy age cannot be smaller than the dynamical timescale, a lower limit on the SED age is often set. Following \citet{wuyts12SED} we use 70 Myr as a lower limit, and the age of the Universe at the observed redshift as an upper limit. The effect of the age limit is not critical: lowering it to 20 Myr causes an average increase of 0.10 dex in both stellar mass and star formation rate.

One of the parameters involved in the SED fitting is the metallicity of the stellar population. This is different from the gas-phase metallicity, that we measure from rest-frame optical emission lines (see Section \ref{sec:metallicity}) and that can be higher than the stellar metallicity. The allowed values of stellar metallicity for the SED fitting are 0.2, 0.4 and 1.0 in units of solar metallicity. For some of the arcs, the best-fit value is significantly larger than the gas-phase metallicity. We attribute this unphysical result to the effect of SED fitting degeneracies. We performed a test for each arc by fitting the SED while keeping the metallicity fixed at the value closest to the one measured via emission lines. This results in slightly larger values of dust extinction, which is degenerate with metallicity. However, the effect is small: the offset is nearly always smaller than the error bar and the average change in dust extinction is $\langle \Delta E(B-V) \rangle = 0.07 $. The other stellar population parameters are negligibly affected, and their uncertainties change only marginally.

It is important to note that among the SED fitting output parameters, stellar mass is the most robust \citep[e.g.][]{wuyts07}, and is also the only one that is critical for our analysis. Star formation rate and dust extinction are more sensitive to the assumptions made, but it is possible to compare them with independent measurements from the rest-frame optical emission lines.

\begin{deluxetable*}{llllllll}
\tabletypesize{\footnotesize}
\tablewidth{0pc}
\tablecaption{Emission Line Fluxes \label{tab:lines}}
\tablehead{
\colhead {Source} & \colhead{\OII} & \colhead{\NeIII} & \colhead{\Hbeta} & \colhead{\OIIIa} & \colhead{\OIIIb} & \colhead{\Halpha} & \colhead{\NII} }
\startdata
A611  		& \alignp \nodata  		& \alignp \nodata		& 15.9 \p 4.5  		& 34.0 \p 8.6\note  	&  90.4 \p 4.1 		& 43.3 \p 6.0  		& \alignnn $<$ 4.1\note \\
RXJ2129  	& \alignp \nodata		& \alignp \nodata		& 10.6 \p 5.1\note  	& 11.8 \p 4.0\note  	&  45.8 \p 7.3  	& 32.8 \p 5.9  		& \alignnn $<$ 3.7    	\\
A1413		& \alignnn $<$ 29\note		& 10.5 \p 3.7	  		& 12.8 \p 3.2\note  	& 60.8 \p 6.3\note  	&  \phantom{.}176 \p 3 	& 97.2 \p 3.7  		& \alignnn $<$ 11\note	\\
A1835  		& \phantom{2.}36 \p 13  	& \Align 5.8 \p 2.1  		& 16.6 \p 3.1  		& \alignnn $<$ 14\note  &  44.3 \p 4.2  	& 45.4 \p 8.1  		& \Align 5.5 \p 2.1    	\\
A773  		& 23.8 \p 9.7  			& \alignnn $<$ 24  		& \Align 8.3 \p 3.5\note& 15.3 \p 3.4  		&  49.0 \p 5.2  	& 47.2 \p 7.6  		& \alignnn $<$ 9.2	\\
MACS0717  	& 19.3 \p 9.0\note  	& \alignnn $<$ 9.1  			& \Align 8.5 \p 2.9  	& \Align 4.4 \p 1.7  	&  18.3 \p 2.3  	& 18.9 \p 4.9  		& \alignnn $<$ 6.0	\\
A383  		& 18.5 \p 5.9  			& \alignnn $<$ 38\note		& \alignnn $<$ 8.5\note & \alignnn $<$ 6.9\note &  16.4 \p 2.2  	& 25.2 \p 7.1  		& \alignnn $<$ 7.4	\\
A1689  		& \alignnn $<$ 29\note  	& \alignnn $<$ 7.7\note 	& \alignnn $<$ 11\note  & \alignnn $<$ 24\note  &  49.1 \p 4.9  	& \alignp \nodata	& \alignp \nodata 	\\
A1703  		& 12.7 \p 5.7  			& \Align 5.5 \p 1.0		& 11.2 \p 2.5  		& 34.0 \p 3.6  		&  81.1 \p 2.0  	& \alignp \nodata	& \alignp \nodata 
\enddata
\tablecomments{Fluxes in units of $10^{-17}$~erg~s$^{-1}$~cm$^{-2}$. The listed uncertainties apply to flux ratios, and do not include the error in the absolute flux calibration. For undetected lines the 2-$\sigma$ upper limit is given.}
\tablenotetext{$\ast$}{Lines strongly contaminated by sky emission lines. The uncertainty on these lines does not include systematic effects due to sky residuals.} 
\end{deluxetable*}

The best-fit spectra are shown in Figure \ref{fig:centerpiece}, and the output parameters (stellar mass, dust extinction, age, and current star formation rate) for each arc are listed in Table \ref{tab:properties}. Stellar masses and star formation rates are corrected for the gravitational magnification. The stellar masses are in the range $7.8 < \log \M / \Msun < 9.4$, and are located at the low end of the mass distribution of the SDSS sample \citep[e.g.][]{zahid12lowmass}. The uncertainties are between 0.1 and 0.3 dex except for A1413, for which the low number of photometric points yields a large uncertainty in the stellar mass, $\Delta \log \M = 0.36$ dex.

It is common practice to report the best-fit stellar mass (i.e.\ the one corresponding to the model that best describes the photometric data) and the 68\% confidence region. The error bars are often highly asymmetric, and are very difficult to propagate when using the SED results in further analysis. In fact, a rigorous propagation of asymmetric error bars is possible only when the posterior distribution is known. Instead, we calculate the stellar mass posterior distribution from the chi-square grid output from FAST, and report the mean and the standard deviation of the distribution. The posterior distributions in $\log \M$ are only weakly skewed and are well approximated by a Gaussian function. On the other hand, the best-fit value is often off-center, and choosing it as the best estimate would cause asymmetric error bars. Reporting the mean and standard deviation has the advantage of a straightforward propagation of the uncertainty in following calculations, which is essential for the present work. The same arguments apply to other stellar population parameters such as log SFR and dust extinction $E(B-V)$, and we follow the same method for estimating their values. The stellar population age, however, presents a posterior distribution that is very skewed for those galaxies with a best-fit age near the lower limit, therefore we list the best-fit value and the asymmetric 68\% confidence interval, which we do not use in any further analysis.


\section{Spectroscopic Diagnostics}
\label{sec:spectra}

The goal of this study is to explore the relation between stellar mass, star formation rate and gas metallicity for star-forming galaxies at high redshift. In the previous section we derived the stellar masses using photometric data. In this section we use the rest-frame optical emission lines of the gravitational arcs to measure their star formation rate and metallicity.

In order to derive physical quantities from the observed spectra, we need to quantitatively analyze the emission lines. Each emission line profile was fitted with one Gaussian (two for the doublet \OII). For each line we derived flux, redshift, width and continuum level from the fit. For the faintest lines we fixed one or more of these parameters using as a reference \OIIIb\ or \Halpha, which have a relatively high signal to noise ratio. The line fluxes are reported in Table \ref{tab:lines}. Some of the emission lines fall in the vicinity of bright sky emission features. In such cases, sky subtraction residuals may bias the Gaussian fits, because of an imperfect estimate of the error spectrum. This effect is not included in the random uncertainties given in Table \ref{tab:lines}, however, we mark those measurements which might be affected by a large systematic error.

The Galactic extinction is very small for all the objects considered, and negligible compared to the uncertainty on the fluxes.

The redshifts of the gravitational arcs, measured from \OIIIb, are given in Table \ref{tab:sample}. In two cases we found that previously published redshifts were incorrect \citep[A611 and RXJ2129,][]{richard10} due to misidentification of rest-frame UV spectral features.

\begin{figure}[bp]
\centering
\includegraphics[width=0.45\textwidth]{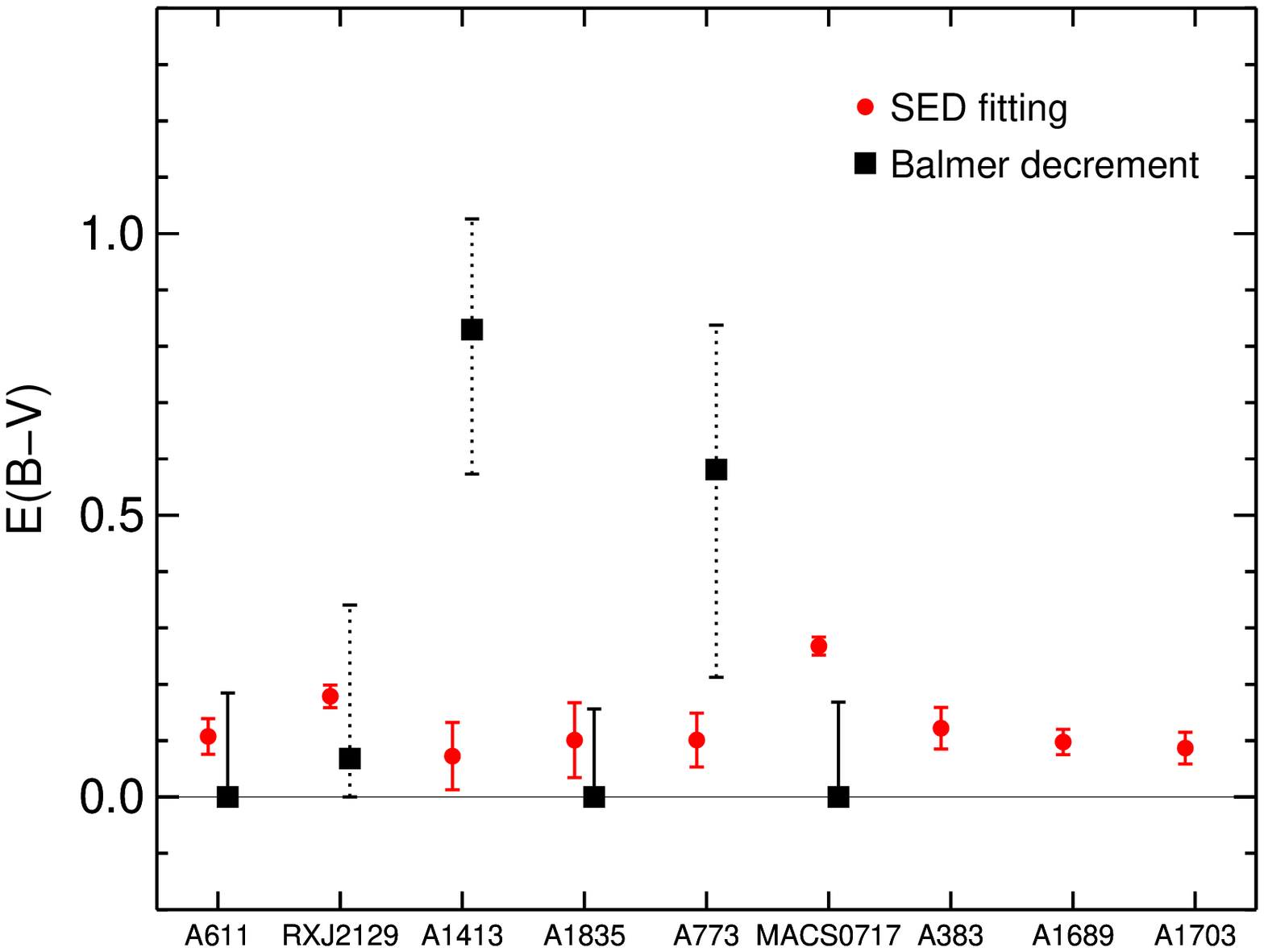}
\caption{Dust extinction $E(B-V)$ derived from SED fitting (red circles) and Balmer decrement (black squares). Dotted error bars indicate measures affected by sky line contamination.}
\label{fig:dust}
\end{figure}

\begin{deluxetable*}{ccccccccc}
\tabletypesize{\footnotesize}
\tablewidth{0pc}
\tablecaption{Physical Properties of the Sample \label{tab:properties}}
\tablehead{
\colhead {Source}  & \colhead{log(\M/\Msun)\tablenotemark{a}} & \colhead{$E(B-V)$\tablenotemark{a}} & \colhead{log(Age/yr)\tablenotemark{a}} & \colhead{SFR$_{\rm SED}$\tablenotemark{a}} & \colhead{SFR$_\Halpha$\tablenotemark{b}} & \colhead{Line width\tablenotemark{b}} & \colhead{12 + \logOH\tablenotemark{b}}  & \colhead{FMR residual\tablenotemark{c}}
\\
 &  &  &  & (\Msun/yr) & (\Msun/yr) & (km/s) & & 
} 
\startdata
A611  &  8.27 \p 0.09  &  0.11 \p 0.03  &  $ 8.3 ^{ + 0.3 } _{ - 0.4} $  &  1.7 \p 0.5  &  2.0 \p 1.1  		& 27 \p 5	&  7.89 \p 0.19  &  -0.11 \p 0.20  \\ 
RXJ2129  &  7.80 \p 0.25  &  0.18 \p 0.02  &  $ 7.9 ^{ + 0.1 } _{ - 0.0} $  &  0.9 \p 0.5  &  0.6 \p 0.4  	& \phantom{2}39 \p 10	&  7.89 \p 0.40  &  \phantom{-}0.04 \p 0.42  \\ 
A1413  &  8.72 \p 0.36  &  0.07 \p 0.06  &  $ 8.0 ^{ + 0.5 } _{ - 0.1} $  &  3.1 \p 1.7  &  7.2 \p 4.3  	& 30 \p 3	&  7.89 \p 0.33  &  -0.26 \p 0.38  \\ 
A1835  &  8.33 \p 0.22  &  0.10 \p 0.07  &  $ 8.8 ^{ + 0.4 } _{ - 1.0} $  &  1.1 \p 0.7  &  1.0 \p 0.7  	& 54 \p 7	&  8.45 \p 0.07  &  \phantom{-}0.37 \p 0.14  \\ 
A773  &  9.16 \p 0.21  &  0.10 \p 0.05  &  $ 8.7 ^{ + 0.4 } _{ - 0.5} $  &  4.6 \p 2.7  &  4.4 \p 2.7  		& 50 \p 7	&  8.32 \p 0.11  &  -0.09 \p 0.17  \\ 
MACS0717  &  9.36 \p 0.19  &  0.27 \p 0.02  &  $ 7.9 ^{ + 0.1 } _{ - 0.1} $  &  34 \p 15  &  15 \p 10	  	& 65 \p 8	&  8.53 \p 0.10  &  \phantom{-}0.10 \p 0.15  \\ 
A383  &  8.67 \p 0.16  &  0.12 \p 0.04  &  $ 8.1 ^{ + 0.6 } _{ - 0.3} $  &  3.4 \p 1.3  &  3.8 \p 2.3  		& 54 \p 8	&  8.56 \p 0.10  &  \phantom{-}0.40 \p 0.14  \\ 
A1689  &  8.27 \p 0.24  &  0.10 \p 0.02  &  $ 8.0 ^{ + 0.3 } _{ - 0.2} $  &  2.2 \p 1.3  &   \nodata  		& 80 \p 9	&  7.89 \p 0.39  &  -0.10 \p 0.41  \\ 
A1703  &  8.49 \p 0.27  &  0.09 \p 0.03  &  $ 8.6 ^{ + 0.7 } _{ - 0.7} $  &  1.0 \p 0.5  &  4.1 \p 2.9  	& 34 \p 3	&  7.84 \p 0.14  &  -0.23 \p 0.21  
\enddata
\tablecomments{Stellar masses and star formation rates are corrected for the gravitational magnification.}
\tablenotetext{a}{Derived from SED fitting.}
\tablenotetext{b}{Derived from rest-frame optical emission lines.}
\tablenotetext{c}{Difference between the measured metallicity and the metallicity predicted by the FMR as formulated by \citet{mannucci11}.}

\end{deluxetable*}

\subsection{Dust Extinction}
\label{sec:dust}

We estimated dust reddening using the \Halpha/\Hbeta\ flux ratio. Assuming a case B recombination and typical temperature (10,000 K) and density (100 cm$^{-3}$), the theoretical value of the ratio is 2.87 \citep{osterbrock06}. We used the \citet{calzetti00} law to derive the dust extinction from the observed flux ratio. Since a negative extinction is unphysical, but can be consistent with the measured Balmer decrement because of large uncertainties, we take a Bayesian approach and use a flat, positive prior for $E(B-V)$. The results are shown in Figure \ref{fig:dust} and compared to the dust extinction derived from the SED fitting. The dotted error bars indicate the measurements affected by sky emission. The two methods are in good agreement and provide very low dust extinction for most of the arcs. The only object with a large discrepancy between the two measurements of dust extinction is A1413, for which the Balmer decrement gives $E(B-V) \sim 0.8$. We attribute this very large value to the effect of sky emission on the \Hbeta\ flux, since all the other galaxies in our sample have $E(B-V)<0.3$. The SED fit for this galaxy, although uncertain, being based on only two photometric points, gives a dust extinction very similar to that found for the other arcs, $\langle E(B-V) \rangle = 0.13$.

The Balmer decrement probes the extinction in the HII regions, where the nebular emission originates, while the SED fit output applies to the overall stellar population of a galaxy. In principle by comparing the dust extinction obtained by the two methods it is possible to study the dust distribution, which can be concentrated in star-forming regions. \citet{calzetti94} found that for local starburst galaxies the nebular dust extinction is roughly twice the extinction of the stellar continuum. At $z \sim 2$ it is not clear whether there is a difference between the reddening experienced by stellar and gas emission \citep[e.g.][]{erb06c,hainline09,forsterschreiber09}. Our data suggest a similar amount of attenuation for the two components, but the \Hbeta\ flux determinations are too noisy to draw any conclusion. When in the following analysis we correct the emission line fluxes for dust extinction, we always use the SED fitting values, which are less affected by uncertainty. This method could underestimate the dust extinction experienced by gas emission by a factor of 2, which translates into an average increase in SFR of only 0.15 dex.

\subsection{Lack of AGN Contribution}

\begin{figure}[tbp]
\centering
\vspace*{0.02\textwidth} 
\includegraphics[width=0.45\textwidth]{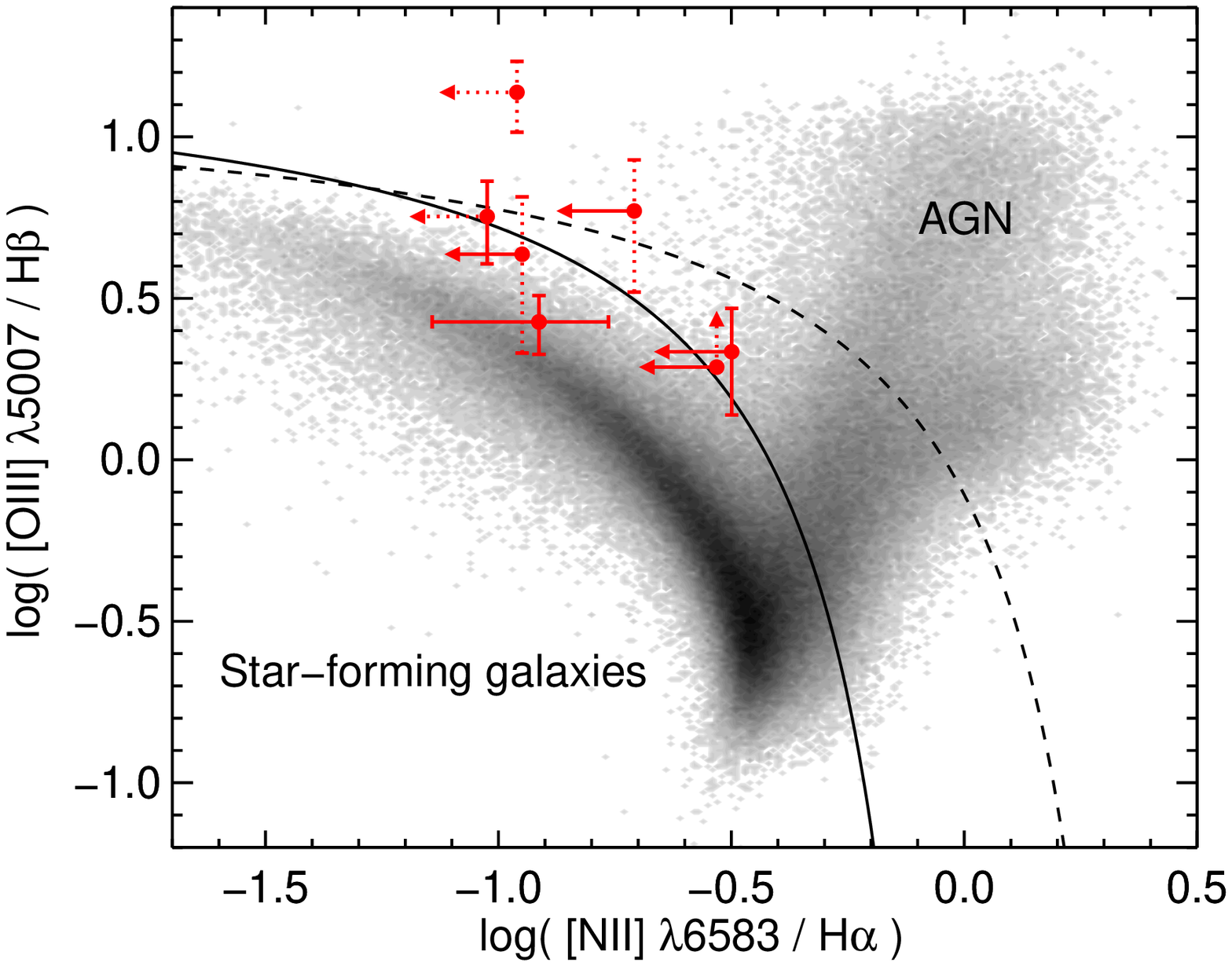}
\caption{BPT diagram \citep{baldwin81}. Our sample (red points) is compared to the SDSS sample \citep[gray density map,][]{kauffmann03}. Also shown are the theoretical \citep[dashed line,][]{kewley01} and empirical \citep[solid line,][]{kauffmann03} separation between active galactic nuclei and star-forming galaxies. Dotted error bars indicate the line ratios that are contaminated by sky emission.}
\label{fig:bpt}
\end{figure}

To exclude the possibility of any AGN contribution to the gravitational arc emission, in Figure \ref{fig:bpt} we show the \OIIIb/\Hbeta\ vs. \NII/\Halpha\ line ratio diagram \citep[BPT diagram,][]{baldwin81}. In this plot star-forming galaxies and AGN populate separate regions due to the different ionization mechanisms at the origin of the line emission. All the gravitational arcs lie on or near the star-forming branch of the diagram, and we can firmly exclude the presence of AGN in our sample. 

It is interesting to note that the location of these $z \sim 2$ star-forming galaxies on the BPT diagram is not coincident with the region most populated by low-redshift galaxies. In particular, none of our objects lie in the region $\mathrm{log}( \OIIIb / \Hbeta) < 0$, where the majority of local galaxies are found. A large \OIIIb / \Hbeta\ ratio has already been reported in many studies of high-redshift star-forming galaxies \citep{shapley05metal, erb06a, erb10, hainline09, richard11, rigby11}, and is indicative of a high ionization parameter, as extensively discussed by \citet{erb10}.

\subsection{Metallicities}
\label{sec:metallicity}

Rest-frame optical nebular lines contain a large amount of information on the physical conditions of the gas responsible for the emission, including its metallicity. If the auroral line [OIII]$\lambda4363$ is detected, then it is possible to calculate the electron temperature and have a direct measurement of the metallicity. Unfortunately this line is so weak that at high redshift it has been detected only for a handful of objects. Instead, we derived the gas metallicity from the flux ratio of strong emission lines. There are several well-established methods to estimate the metallicity from flux ratios, calibrated using either theoretical calculations or observations of low-redshift galaxies. The absolute metallicity obtained with these methods is highly uncertain, and the different sets of calibrations, when applied to the same observations, give results that can differ by as much as 0.7 dex \citep{kewley08}. Although this discrepancy makes it very difficult to compare observational results obtained with different calibrations, relative measurements obtained with the same strong line method are much more reliable.

The main goal of the present study is to test whether the locally determined fundamental metallicity relation applies to high-redshift galaxies as well. The natural choice is then to use the same metallicity calibrations adopted by \citet{mannucci10} in the definition of the local relation. These are the empirical calibrations of \citet{maiolino08}, which give a polynomial fit for the value of various nebular line ratios as a function of the gas metallicity. The main line ratios are \OIIIb/\Hbeta\ and \OIIIb/\OII, while \NII/\Halpha\ and \NeIII/\OII\ were used only for some arcs, mostly as upper or lower limits. We also used \OIIIb/\Halpha, whose calibration we derive from \OIIIb/\Hbeta\ assuming the theoretical value for the Balmer ratio. We note, however, that these two line ratios do not give independent measurements of the metallicity. From the plots shown in \citet{maiolino08} we estimate a scatter in the relations between line ratios and abundance of 0.10 dex, and we add this contribution to the uncertainty calculation.

\begin{figure}[tbp]
\centering
\includegraphics[width=0.45\textwidth]{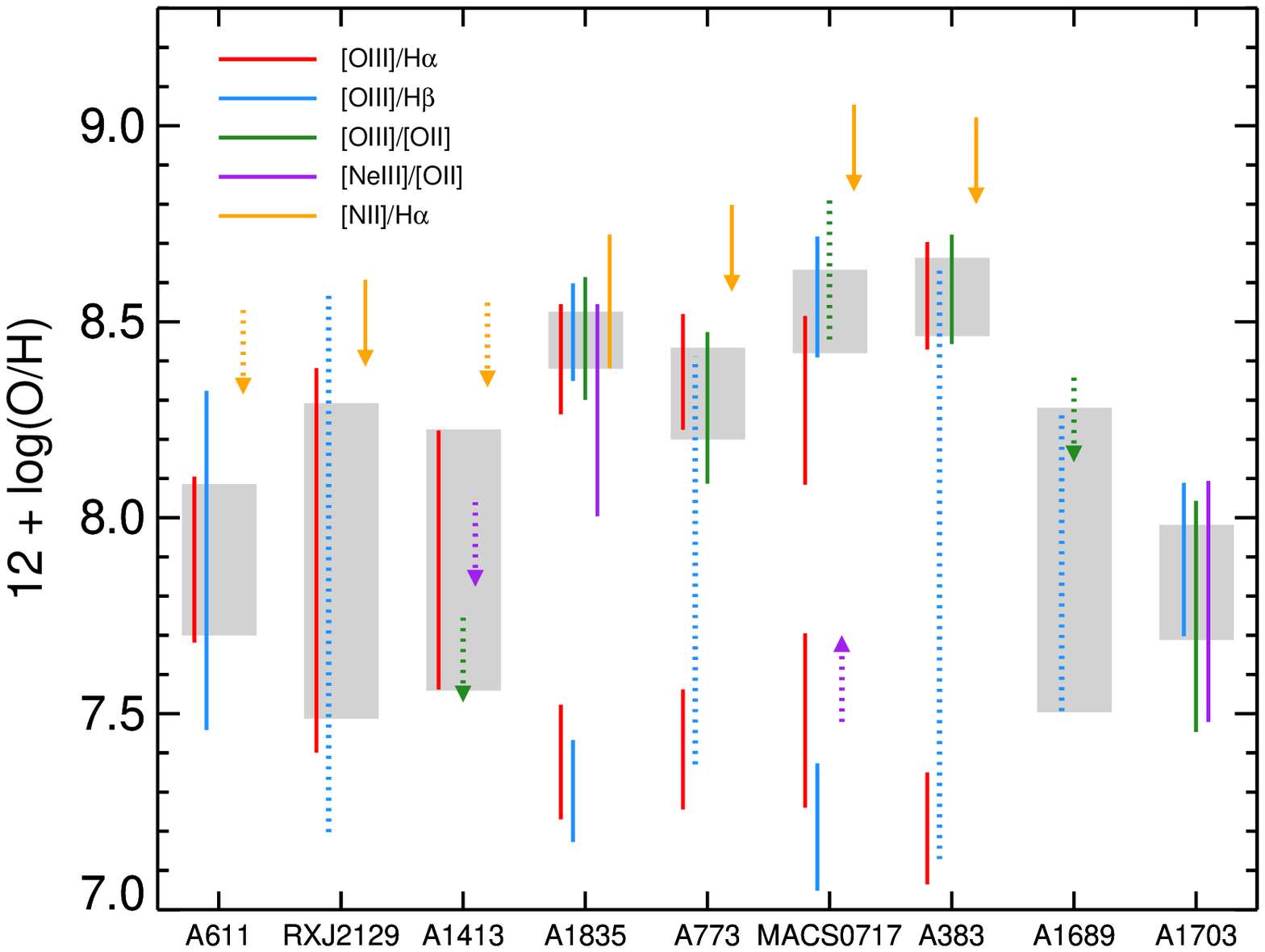}
\caption{Gas metallicity derived using different line ratios adopting the calibrations of \citet{maiolino08}. For each object the gray region shows the weighted mean. Dotted lines indicate ratios involving at least one line contaminated by sky emission.}
\label{metallicities}
\end{figure}

Figure \ref{metallicities} shows the gas metallicity for our sample, derived using the available line ratios. For each arc the final metallicity is the weighted average of the single measurements, not considering upper or lower limits, and is shown in gray in the plot (and listed in Table \ref{tab:properties}). From this figure it is clear that the different line ratios give results always consistent within the error bars, and this is an important confirmation of the reliability of this method.

The relation between \OIIIb/\Hbeta\ and the metallicity is not monotonic, and presents a maximum at $12 + \logOH = 7.89$. Since this is a stationary point, any uncertainty in the line ratio is transformed into a much larger uncertainty in the metallicity. About half of our sample is found in this location, with large uncertainties on \logOH, up to 0.4 dex. The shape of this calibration also causes the existence of two possible metallicity values in some cases, but an unambiguous solution is always found thanks to the other line ratios.

In Figure \ref{metallicities} the metallicities derived from diagnostics that involve at least one line contaminated by sky emission are plotted as dotted lines. They are generally consistent with the other line ratios, although most of them have very large uncertainties and do not influence the weighted average in an appreciable way. We therefore conclude that our results do not depend on the emission lines affected by sky residuals. The only exception is A1689, for which only one upper limit and one lower limit on the line ratios are available, and both may be contaminated by sky emission. Although the lower limit on the \OIIIb/\Hbeta\ ratio gives a finite confidence interval in \logOH\ thanks to the non-monotonic metallicity calibration, we note that the abundance for this galaxy is not reliable.

One of the most widely used metallicity diagnostics is $R_{23}$, defined as the ratio between the oxygen lines $(\OII + \OIIIa + \OIIIb)$ and \Hbeta. Although this line ratio, with the \citet{maiolino08} calibration, gives  results that are consistent with the other diagnostics, we do not use it because it is not independent on the line ratios \OIIIb/\Hbeta\ and \OIIIb/\OII. Using these two line ratios instead of $R_{23}$ has the advantage of isolating the abundance determinations which are affected by sky residuals.

It is worth remarking that the ratio between the flux of two lines is independent of the absolute flux calibration even for lines that lie in very distant parts of the spectrum, since Triplespec allows us to observe the $J$, $H$ and $K$ band simultaneously. It is also independent of gravitational magnification and slit loss. We corrected the line fluxes for dust extinction, using the SED fitting results, in a differential way: the ratio of two lines depends only on the ratio of the attenuation at the corresponding wavelengths. This results in a small correction to the metallicity estimate and its uncertainty. 

The metallicity of the two gravitational arcs A1689 and A1835 has already been measured by \citet{richard11} from near-infrared spectra obtained with Keck NIRSPEC and using the same set of \citet{maiolino08} calibrations. Our results are in good agreement for both arcs. In particular, \citeauthor{richard11} detect \Hbeta\ and \OIIIb\ in the spectrum of A1689, obtaining  $12 + \logOH = 8.00^{+0.44}_{-0.50}$, a value very close to our estimate. For this reason we will not exclude A1689 from our sample despite the poor quality of its spectrum.

\subsection{Star Formation Rates}

\begin{figure}[tbp]
\centering
\includegraphics[width=0.48\textwidth]{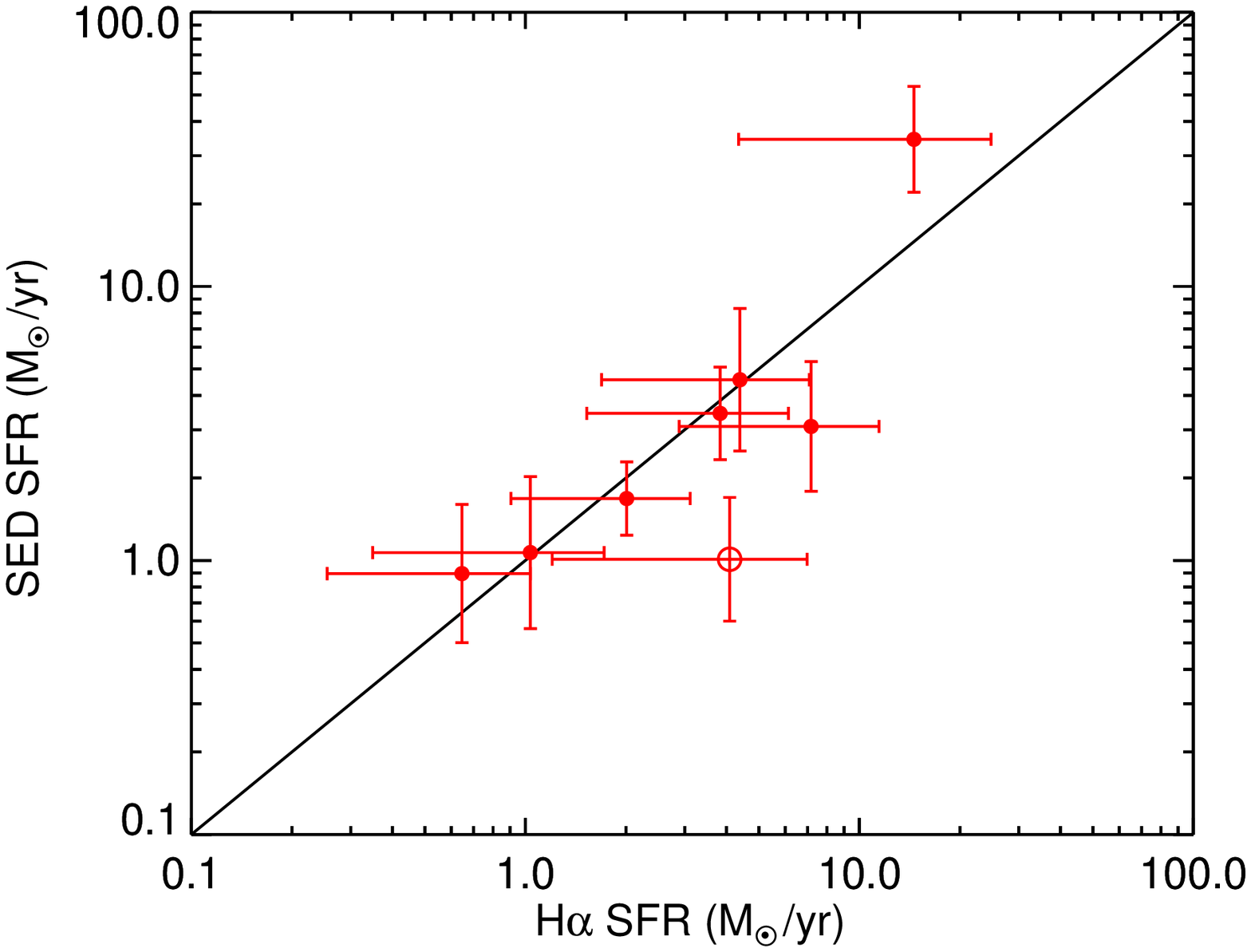}
\caption{Comparison of star formation rates derived using two methods: SED-fitting and \Halpha\ flux. The empty circle is A1703 for which \Hbeta\ has been used as a proxy for \Halpha.}
\label{comparison_SFR}
\end{figure}

We derived the current star formation rate from the extinction-corrected \Halpha\ emission flux using the calibration given by \citet{kennicutt98}, dividing the result by 1.7 to convert to that appropriate for a Chabrier IMF. The resulting star formation rates, corrected for the gravitational magnification, are reported in Table \ref{tab:properties}.

The SFR derived from nebular emission accounts for the star formation activity in the physical region of the arc that is covered by the slit, which is different, in principle, from the star formation rate of the entire galaxy. But the narrow gravitational arcs from our sample are easily covered by the 1 arcsec wide slit, as is apparent from the image stamps in Figure \ref{fig:centerpiece}. Therefore we do not attempt to correct for this effect, which is in any case less important than the uncertainty caused by slit alignment and seeing variability.

Figure \ref{comparison_SFR} shows excellent agreement between the star formation rates calculated using SED fitting and \Halpha\ flux. This is encouraging because it validates the numerous assumptions made in the derivation of the star formation rates. It is particularly interesting that the agreement between SED-fitting and nebular emission even holds for A1413, where only two photometric points are available. It is possible that the simplifying choice of a constant star formation history, with the consequent decrease in the number of free parameters, helped reduce the scatter in the comparison between the two methods.

In the following section we will always use the star formation rate derived from the \Halpha\ flux. The spectra of the two objects at $z>3$ do not include \Halpha, which is redshifted outside the Triplespec range. For one of them (A1703, empty circle in Figure \ref{comparison_SFR}) we use the observed flux of \Hbeta\ as a proxy for \Halpha, assuming the theoretical line ratio discussed in Section \ref{sec:dust} and correcting for dust extinction. For A1689, for which both \Halpha\ and \Hbeta\ are not available, we use the SED fitting star formation rate.

\subsection{Line Widths}

The broadening of the emission lines due to the gas kinematics depends on the gravitational well of the galaxies. Measuring the line widths can then give an estimate on the gravitational arc masses that is independent of SED fitting and gravitational lensing models.

The observed velocity width of the nebular lines need to be corrected for the instrumental resolution, which was measured from the sky OH lines. For each source we extracted the spectrum of the sky using the same procedure followed to extract the arc spectrum. We measured the dispersion of the brightest, unblended OH lines, which is 40-55 km s$^{-1}$ depending on spectral order and wavelength. We calculated a linear fit of the ratio of the spectral resolution $R$ to the order $m$ as a function of wavelength, and used this to estimate the instrumental resolution for each nebular line.

Most of the arc emission lines are well-resolved. For each source we take the weighted mean of the line widths of all the well-detected lines excluding \OII, which is a doublet and is not completely resolved. The results are listed in Table \ref{tab:properties}.

Since a detailed lensing map is needed to measure the intrinsic radius of the gravitational arcs, we do not attempt to estimate the dynamical masses. The observed velocity dispersions, however, are unusually low if compared to the results of similar studies \citep{law09, forsterschreiber09, jones10}. This is an important confirmation of the low masses found in our sample.


\section{Results}

In this section we combine our measurements of stellar mass, gas metallicity and star formation rate to explore the properties of our sample of low-mass galaxies. In particular, the goal of this work is to test whether high-redshift galaxies follow the local fundamental metallicity relation, claimed to be valid up to at least $z \sim 2.2$ \citep{mannucci10}.

\subsection{The Mass--Metallicity Relation}
\label{sec:mzr}

\begin{figure}[tbp]
\centering
\includegraphics[width=0.45\textwidth]{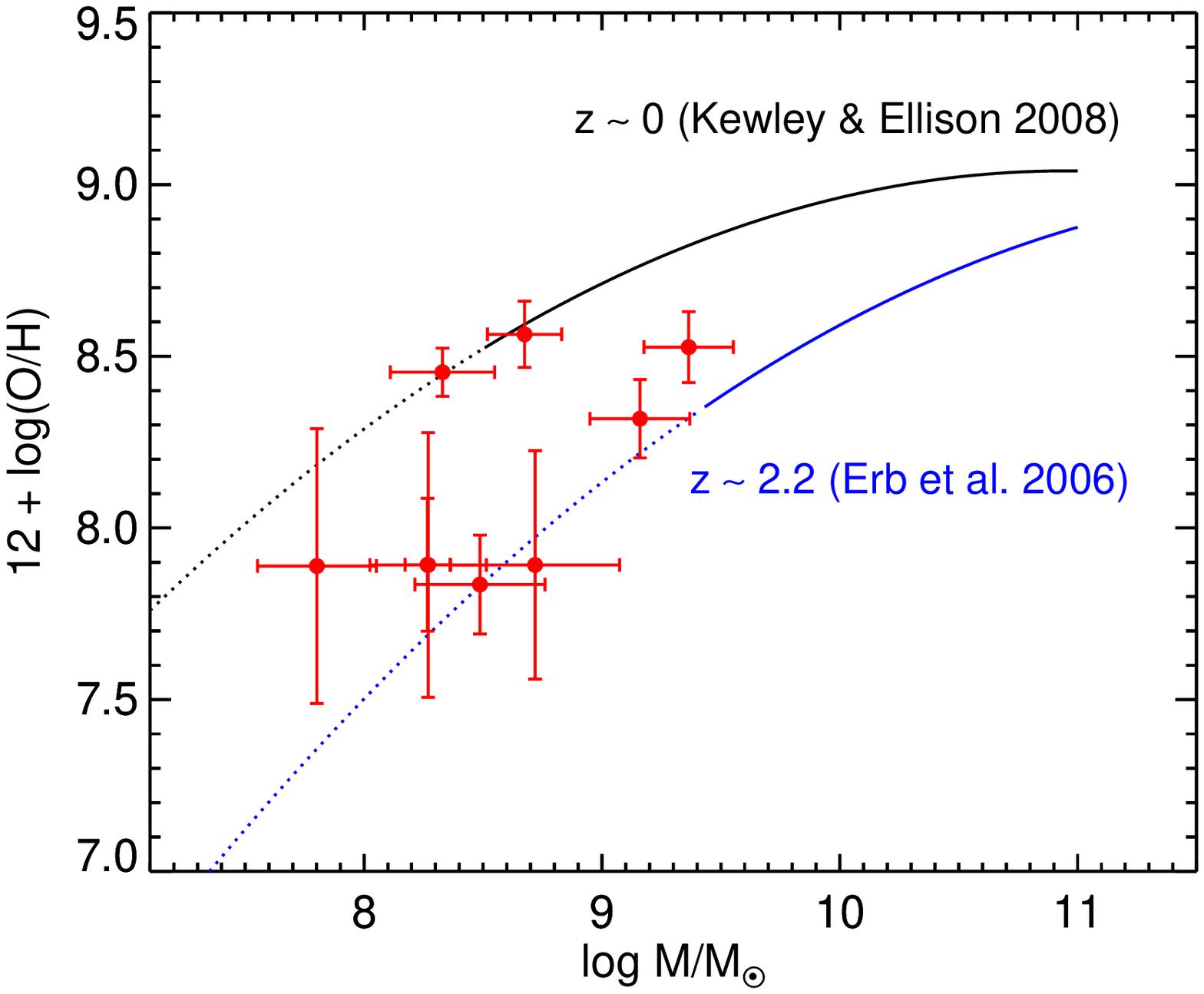}
\caption{Mass-metallicity relation for our sample (red points). The fit to the relation at $z \sim 0$ \citep[black solid line,][]{kewley08} and $z \sim 2.2 $ \citep[blue solid line,][]{erb06a} are also shown. Note that these fits are calculated using the \citet{maiolino08} metallicity calibrations. The dashed lines are extrapolation at low masses.}
\label{massmet}
\end{figure}

\begin{figure}[tbp]
\centering
\includegraphics[width=0.45\textwidth]{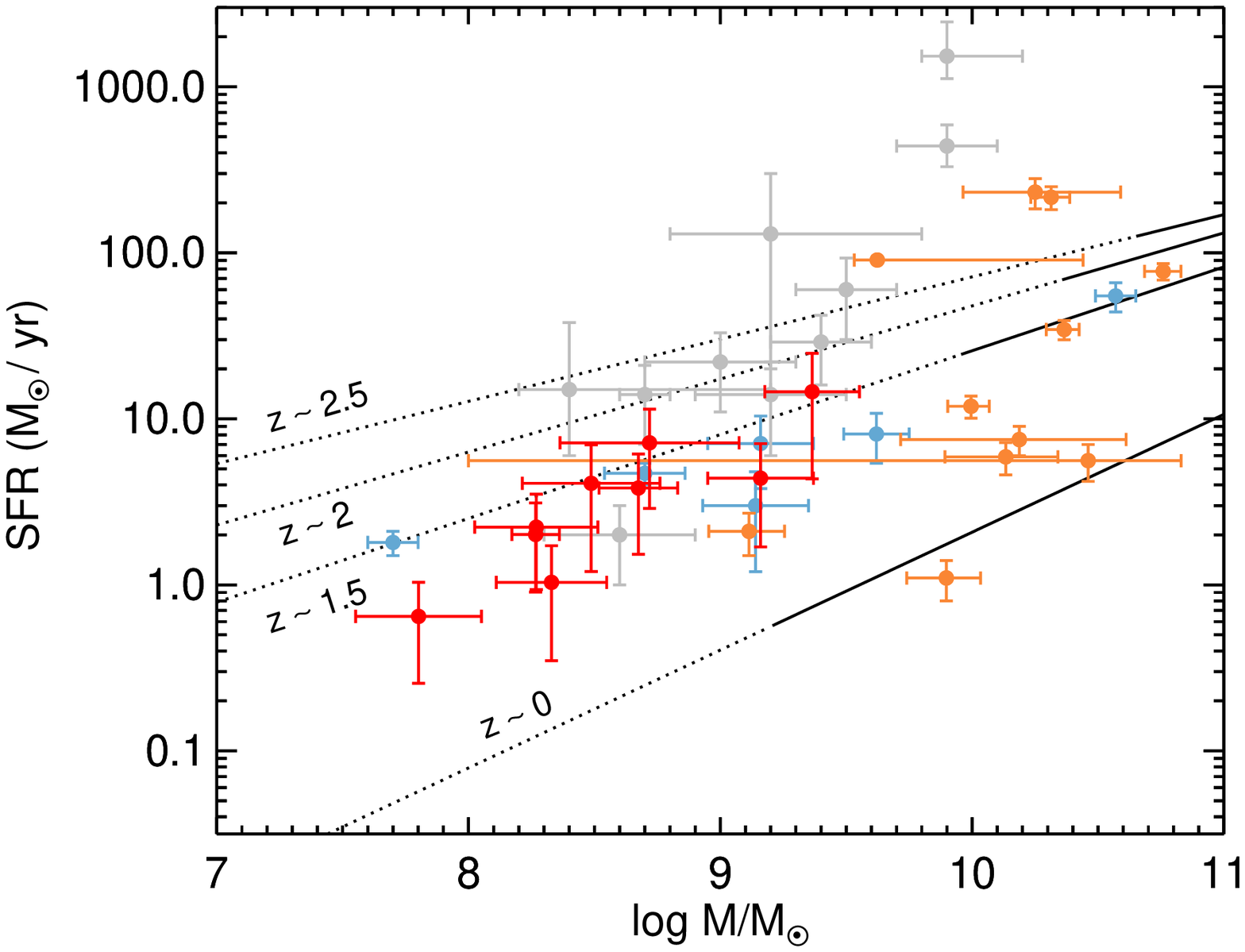}
\caption{Star formation rate versus stellar mass diagram. The main sequence of star-forming galaxies is shown in black at $z \sim 0$ \citep{zahid12model} and at high redshift \citep{whitaker12}, with dotted lines being the extrapolation at masses below the completeness limit. Our sample (red points) is roughly on the main sequence at $z \sim 2$. Points from other studies of lensed galaxies at high redshift are shown: \citet{richard11} in orange, \citet{wuyts12} in gray, and \citet{christensen12} in blue.}
\label{mainseq}
\end{figure}

The mass-metallicity relation for our sample is plotted in Figure~\ref{massmet} together with the fit to the local relation from \citet{kewley08} and to the high-redshift one of \citet{erb06a}. All the results shown in this figure have been derived using the same set of metallicity calibrations \citep[the fits shown are taken from][and are corrected for the choice of IMF]{maiolino08}.

The magnification caused by gravitational lensing allows us to probe stellar masses much smaller than those considered in previous studies of unlensed galaxies, even in the nearby Universe. This makes a direct comparison difficult, but it is clear from Figure \ref{massmet} that our points do not lie on the extrapolation of neither the $z \sim 0$ nor the $z \sim 2$ relations, and have a substantial scatter, larger than the observational uncertainties. 

\citet{yuan13} measured the mass-metallicity relation for a sample of gravitational arcs at $z \sim 2$ obtained by combining their data with results from previous studies. Although a comparison of the absolute measurements is not possible because they use a different metallicity calibration, their results are similar to ours: the lensed galaxies tend to have lower metallicities than the SDSS galaxies but do not lie on a tight sequence.

This results are consistent with the hypothesis of a mass-metallicity relation dependent on some other parameter that is not necessarily the redshift. In the remaining parts of this section we will explore the role of the star formation rate.

\subsection{Star Formation Rate versus Stellar Mass}

In Figure~\ref{mainseq} we show the location of our sample in the SFR-stellar mass diagram (red points), compared to the results of other studies of lensed galaxies at high redshift: \citet[][in orange]{richard11}, \citet[][in gray]{wuyts12}, and \citet[][in blue]{christensen12}. Our sample populates the lower left corner, with masses and star formation rates on average lower than what probed by previous studies. In particular, we more than doubled the number of low-mass galaxies ($\M < 10^9 \Msun$) at this redshift with known metallicity and SFR.

In the mass-SFR plane, star-forming galaxies lie on a relatively tight relation often called the main sequence \citep{noeske07}. This sequence evolves strongly with redshift, with the normalization decreasing over cosmic time at least since $z \sim 2.5$ \citep[][black lines in Figure \ref{mainseq}]{whitaker12, zahid12model}. The gravitational arcs that we selected have star formation rates that are on or below the main sequence at $z \sim 2$. Previous studies of lensed galaxies did not reach such low values of SFR, with the exception of the work of \citet{christensen12}. 

The very low star formation rate of these arcs, between 0.6 and 15 \Msun/yr, is of fundamental importance in this study. First,  the fact that our sample lies on the main sequence means that these galaxies are representative of the typical population of star-forming galaxies. Shallower studies are biased towards luminous galaxies with star formation rates much higher than the main sequence. These objects are thought to be in a starburst phase, potentially caused by a merger, and are not representative of the typical conditions of star-forming galaxies. Secondly, it allows us to compare high and low redshift galaxies with similar star formation rates, and thereby directly address the goals of this paper. From Figure \ref{mainseq} we can see that local massive galaxies ($9.5 < \log \M/\Msun < 11 $) that lie on the main sequence at $z \sim 0$ have SFRs comparable to our sample. This is not the case for the majority of lensed galaxies considered by previous studies.

\subsection{The Fundamental Metallicity Relation}
\label{sec:fmr}

\begin{figure*}[tbp]
\centering
\includegraphics[width=1\textwidth]{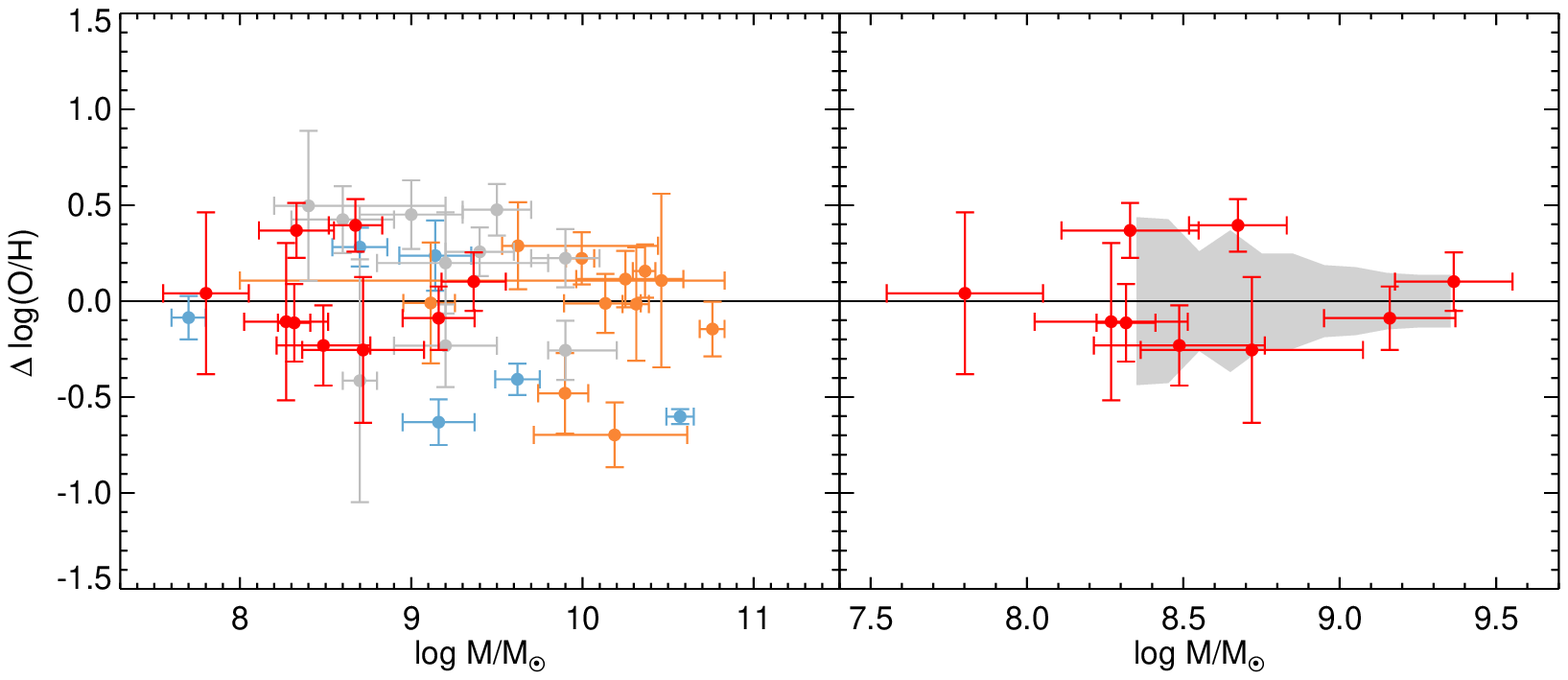}
\caption{Difference between measured metallicity and the prediction of the fundamental metallicity relation. \emph{Left:} Our residuals (red points) are compared to previous high-redshift studies, color-coded as in Figure \ref{mainseq}. \emph{Right:} Our points are compared to the SDSS low-mass sample, whose standard deviation is shown in gray. In both panels A611 is offset by 0.05 dex in mass for clarity.}
\label{FMR}
\end{figure*}

We now turn our attention to the fundamental metallicity relation, a surface in the 3D parameter space of stellar mass, metallicity and star formation rate tightly followed by the SDSS galaxies discovered by \citet{mannucci10} and extended to low masses by \citet{mannucci11}. In Figure \ref{FMR} we plot for each gravitational arc the difference between the metallicity that we measure from nebular lines and the metallicity predicted by the local FMR given its stellar mass and star formation rate. We also show the points from previous studies of lensed galaxies. Although high-redshift galaxies seem to roughly follow the local relation, some of the samples shown in Figure \ref{FMR} show a systematic offset. Since our sample is more strictly selected in terms of star formation, we will limit the quantitative analysis to our 9 gravitational arcs.

The weighted average of the residuals for our sample is $0.12 \p 0.06$ dex. However, the weighted mean is skewed toward galaxies with higher metallicity since they are measured with higher precision, because the metallicity calibrations are not linear. This is clear from Figure \ref{metallicities}, where the group of galaxies aligned at $12 + \logOH \sim 7.9$, the maximum of the \OIIIb/\Hbeta\ line ratio calibration, present the largest error bars. The arithmetic mean is not affected by this bias, and gives $\langle \Delta \logOH \rangle = 0.01 \p 0.08$. The agreement of high-redshift lensed galaxies with the local fundamental metallicity relation is remarkable, and strongly suggests that these objects lie on the relation independently of their redshift at least up to $z \sim 3$. This represents the first clear result for high-redshift galaxies with $M < 10^9 \Msun$, and the first time that the universality of the FMR is confirmed using galaxies at high redshift with a SFR which is observed in typical galaxies in the local Universe.

\subsection{The Scatter in the Fundamental Metallicity Relation}
\label{sec:fmr_sigma}

Our gravitational arcs show a relatively small scatter around the local fundamental metallicity relation. The standard deviation of the metallicity offsets from the FMR is 0.24 dex, and the mean error in $\Delta\logOH$ is 0.25 dex. Also, none of the gravitational arcs is more than $3\sigma$ away from the local fundamental metallicity relation. These two facts suggest that the observed scatter could be in principle just a product of observational uncertainties. In contrast, the standard deviation found by \citet{mannucci11} for the SDSS sample is about 0.4 dex at $10^{8.4}\Msun$, and is shown in gray in the right panel of Figure \ref{FMR} as a function of stellar mass. Note that roughly $32\%$ of the galaxies in the local sample fall outside of the shaded area, while only one among the high-redshift galaxies does not lie in this region. Although \citet{mannucci11} do not report the typical errors on mass, star formation rate and metallicity, they claim that the observational uncertainties are not large enough to explain the observed dispersion. Furthermore, SDSS galaxies are selected by requiring a signal to noise ratio greater than 25 for the \Halpha\ flux, therefore the uncertainty in their metallicity must be much smaller than for our sample.

In order to facilitate the comparison with studies at different redshifts, we estimate the intrinsic scatter in the fundamental metallicity relation using a Bayesian framework. We assume that each measured metallicity residual $\Delta_i \equiv \Delta \logOH_i $ is normally distributed around its \emph{true} value $\tilde \Delta_i$ with standard deviation given by the observational uncertainty $\sigma_i$:
\begin{equation}
	p( \Delta_i | \tilde \Delta_i , \sigma_i ) = \frac{ 1 }{ \sqrt{ 2 \, \pi \, \sigma_i^2 } } \, \exp \left[ -\frac{1}{2} \, \frac{ \left( \Delta_i - \tilde \Delta_i \right)^2 }{ \sigma_i^2 } \right] \; .
\end{equation}
We also assume that the true values $\tilde \Delta_i$ are normally distributed around zero with an intrinsic dispersion $\tilde \sigma$:
\begin{equation}
	p( \tilde \Delta_i | \tilde \sigma ) = \frac{ 1 }{ \sqrt{ 2 \, \pi \, \tilde \sigma^2 } } \, \exp \left[ -\frac{1}{2} \, \frac{ \tilde \Delta_i^2 }{ \tilde \sigma^2 } \right] \; .
\end{equation}
Note that by centering the Gaussian distribution on zero, we are setting the local FMR to hold at high-redshift. This is in agreement with our observations, as we showed in the previous section.
Since we do not know the true value of each data point, we need to marginalize over $\tilde \Delta_i$ in order to obtain the probability density function of the observed $\Delta_i$:
\begin{equation}
	p( \Delta_i | \sigma_i, \tilde \sigma) = \int \left[ p( \Delta_i | \tilde \Delta_i , \sigma_i ) \cdot p( \tilde \Delta_i | \tilde \sigma ) \right] \, \mathrm{d} \tilde \Delta_i \; ,
\end{equation}
and we finally obtain the likelihood function:
\begin{equation}
\begin{split}
	\mathcal{L} (\tilde \sigma) & = \prod_i p( \Delta_i | \sigma_i, \tilde \sigma) \\
	& = \prod_i \frac{ 1 }{ \sqrt{ 2 \, \pi \, \left( \sigma_i^2 + \tilde \sigma^2 \right) } } \, \exp{ \left( -\frac{1}{2} \, \frac{ \Delta_i^2 }{ \sigma_i^2 + \tilde \sigma^2 } \right) } \; .
\end{split}
\end{equation}
Finally, using a uniform prior, the posterior distribution for $\tilde \sigma$ is simply proportional to the likelihood.

The likelihood function peaks at an intrinsic dispersion of 0.20 dex, and calculating mean and standard deviation gives $\tilde \sigma = 0.24 \pm 0.11 $ dex. This calculation shows that although a zero intrinsic dispersion is very unlikely, our data favor a value smaller than the one found in the local Universe. Despite the low number of data points, we can robustly rule out very large intrinsic dispersions: the $95\%$ confidence interval upper limit is $\tilde \sigma < 0.44 $ dex.


\section{Discussion}

Our data confirm that the fundamental metallicity relation applies to low-mass galaxies at $1.5 < z <3$. This suggests that this relation is time-invariant and therefore universal.

In the equilibrium model of \citet{finlator08}, metallicity and star formation rates are tightly connected to gas inflows and outflows, so that a change in one implies a consequent change in the other \citep[see also][]{dave11II, dave12}. If each of these processes are in equilibrium, then the FMR is naturally explained. If a galaxy is perturbed, by e.g.\ a minor merger, after a certain time it will return to the equilibrium configuration. The observed evolution in the mass-metallicity relation could be due to the fact that we are sampling galaxy populations with different star formation rates at different redshifts. This would explain why our points do not lie on the low-mass end of the $z\sim2$ mass-metallicity relation from \citet{erb06a}, that was determined using relatively high-SFR galaxies. 

The analysis of the FMR scatter may provide a valuable additional constraint for numerical or analytical models of galaxy evolution. This is particularly important at low masses, where models have diverging predictions \citep{zahid12lowmass}. In the equilibrium model, the observed scatter of the FMR is determined by how quickly a perturbed galaxy can return to equilibrium. This timescale, in turn, depends on the mass loading factor, a parameter that is fundamental for hydrodynamic simulations. The observation of the scatter in the FMR at different stellar masses and redshifts therefore gives important constraints on numerical models of galaxy evolution, even though current simulations do not resolve stellar masses below $10^9 \Msun$ \citep[e.g.][]{dave11I}.

Our results suggest a low scatter around the fundamental metallicity relation at high redshift, lower than what found by \citet{mannucci11} in the local Universe. In another study of low-redshift galaxies, \citet{bothwell13} investigated the relation between stellar mass, star formation rate and gas content and found that it is at least as tight as the FMR. Interestingly, they found an increase in the dispersion for $\M < 10^9 \Msun$, a confirmation of the results of \citet{mannucci11}. They attribute the increase in scatter to the fact that low-mass galaxies contain a gas mass comparable to the mass of infalling neutral hydrogen clouds, with the result that the accretion process is not smooth but discontinuous and stochastic.

This trend is also confirmed by the results of \citet{henry13}, which studied a sample of low-mass galaxies at $z =$ 0.6 -- 0.7 and found not only a good agreement with the local fundamental metallicity relation, but also a tight dispersion of 0.20 dex that could be explained by the observational uncertainties.

\citet{hunt12} analyzed a sample of 1100 galaxies at $0<z<3.4$, that includes many low-mass galaxies, and found a fundamental plane in the SFR, stellar mass and metallicity space, which is independent on redshift and with a scatter of 0.17 dex. Although this dispersion is much smaller than the one found locally by other studies, \citet{hunt12} do not investigate the dependence of the scatter on redshift and mass. Most importantly, in contrast to the other studies mentioned so far, they do not use the \citet{maiolino08} metallicity calibrations, so that a direct comparison is very difficult.

An interesting perspective on the issue of the FMR scatter has been pointed out by \citet{zahid12lowmass}. They studied the mass-metallicity relation at $z \sim 0$ using various samples from the literature and found a clear increase in the intrinsic scatter at low stellar masses. They suggest the possibility that this scatter is due to a population of low-mass, metal-rich galaxies which are near the end of their star formation. At a fixed mass, the same amount of metals would give a higher metallicity measurement, since there is little gas left. Assuming that the scatter in the FMR is directly caused by the scatter in the mass-metallicity relation, our observations agree well with this scenario, since at high redshift such a population of low-mass galaxies terminating their star-formation would not be expected.
For conclusive results on the evolution of the intrinsic FMR scatter, however, a larger high-redshift sample is needed, together with a rigorous analysis of the observational uncertainties in the local sample.

\section{Summary}

We present near-infrared spectroscopic data for 9 gravitational arcs between redshift 1.5 and 3.3, and the measurement of their stellar mass, gas metallicity and star formation rate. The use of strong gravitational lensing allows us to probe very low masses and star formation rates. Our sample more than doubles the number of galaxies with stellar masses below $10^9 \Msun$ at $z \sim 2$ with known metallicity and SFR. Our main goal is to test whether these galaxies follow the fundamental metallicity relation discovered for local galaxies.

We find that the gravitational arcs lie above the mass-metallicity relation at $z \sim 2$ but below the local relation. However, they also have SFRs that are roughly on the main sequence of star-forming galaxies. This means that they are representative of typical star-forming galaxies, i.e.\ they are not in a starburst phase. 

Our data are fully consistent with the local fundamental metallicity relation \citep{mannucci10,mannucci11}, with a mean metallicity offset of $0.01 \p 0.08$ dex. The dispersion around the FMR of 0.24 dex is smaller than the one measured for local galaxies, and represents an important additional constraint for galaxy evolution models. \\

We acknowledge Kevin Bundy for providing the WIRC reduction pipeline, Eiichi Egami for the IRAC mosaic images, and Marceau Limousin and Drew Newman for some of the magnification factors.  We thank Drew Newman and Gwen Rudie for useful discussions. We also thank the anonymous referee for helpful comments and suggestions. RSE and SB are supported for this work via NSF grant 0909159. JR is supported by the Marie Curie Career Integration Grant 294074.

\bibliography{tspec}{}

\end{document}